\documentclass[10pt,journal,compsoc]{IEEEtran}

\ifCLASSOPTIONcompsoc
  \usepackage[nocompress]{cite}
\else
  \usepackage{cite}
\fi

\ifCLASSINFOpdf
  \usepackage[pdftex]{graphicx}
  \graphicspath{{./figures/at.vti/}{./figures/miranda/}{./figures/noisyTerrain/}{./figures/viscous/}{./figures/wallmodes/}}
  \DeclareGraphicsExtensions{.pdf,.jpeg,.png}
\else
  \usepackage[dvips]{graphicx}
  \graphicspath{{./figures/at.vti/}{./figures/miranda/}{./figures/noisyTerrain/}{./figures/viscous/}{./figures/wallmodes/}}
  \DeclareGraphicsExtensions{.pdf,.jpeg,.png}
\fi

\usepackage{amsmath}

\usepackage{array}

\ifCLASSOPTIONcompsoc
  \usepackage[caption=false,font=footnotesize,labelfont=sf,textfont=sf]{subfig}
\else
  \usepackage[caption=false,font=footnotesize]{subfig}
\fi

\usepackage{inconsolata}

\usepackage{multicol}
\usepackage{multirow}
\usepackage[table,dvipsnames]{xcolor}
\usepackage{colortbl}

\definecolor{mygray2}{gray}{0.8}
\definecolor{mygray1}{gray}{0.9}
\definecolor{mygray3}{gray}{0.98}

\newenvironment{tttabular}[1]%
{\ttfamily \begin{tabular}{#1}}%
{\end{tabular}}

\usepackage{stfloats}
\usepackage{wrapfig}
\fnbelowfloat

\usepackage{url} 

\usepackage{circuitikz}
\usetikzlibrary{er,positioning,arrows,arrows.meta,patterns,calc,shapes}
\usepackage{ifthen}
\tikzstyle{veryThinEdge} = [line width=0.01cm]
\tikzstyle{thinEdge} = [line width=0.02cm]
\tikzstyle{semiMediumEdge} = [line width=0.03cm]
\tikzstyle{mediumEdge} = [line width=0.04cm]
\tikzstyle{thickEdge} = [line width=0.1cm]
\tikzstyle{dashedEdge} = [dashed, gray]

\tikzstyle{thickNode} = [shape=circle,draw=black,fill=black, inner sep=0pt, minimum size=5]
\tikzstyle{filteredNode} = [shape=circle,draw=filtered,fill=filtered, inner sep=0pt, minimum size=4]

\tikzstyle{arrow} = [-{Stealth[scale=1.5]}]

\tikzset{
  prefix after node/.style={prefix after command=\pgfextra{#1}},
  /semifill/ang/.initial=0,
  /semifill/upper/.initial=none,
  /semifill/lower/.initial=none,
  semifill/.style={
    circle, draw,
    prefix after node={
      \pgfqkeys{/semifill}{#1}
      \path let \p1 = (\tikzlastnode.north), \p2 = (\tikzlastnode.center),
                \n1 = {\y1-\y2} in [radius=\n1]
            (\tikzlastnode.\pgfkeysvalueof{/semifill/ang}) 
            edge[
              draw=none,
              fill=\pgfkeysvalueof{/semifill/upper},
              to path={
                arc[start angle=\pgfkeysvalueof{/semifill/ang}, delta angle=180]
                -- cycle}] ()
            (\tikzlastnode.\pgfkeysvalueof{/semifill/ang}) 
            edge[
              draw=none,
              fill=\pgfkeysvalueof{/semifill/lower},
              to path={
                arc[start angle=\pgfkeysvalueof{/semifill/ang}, delta angle=-180]
                -- cycle}] ();
    }
  }
}

\definecolor{background0_r}{HTML}{f8f2f2}
\definecolor{highlight0_r}{HTML}{e3aeae}
\definecolor{highlight1_r}{HTML}{a93838}

\definecolor{background0_b}{HTML}{f2f6f8}
\definecolor{highlight0_b}{HTML}{b1cfe1}
\definecolor{highlight1_b}{HTML}{3e7fa4}

\definecolor{background0_g}{HTML}{f2f8f5}
\definecolor{highlight0_g}{HTML}{b1e0c3}
\definecolor{highlight1_g}{HTML}{3ea363}

\definecolor{background0_y}{HTML}{f2f6f8}
\definecolor{colorTheme_20}{HTML}{f27d00}
\definecolor{colorTheme_21}{HTML}{f1b473}
\definecolor{highlight0_y}{HTML}{f5ca9c}
\definecolor{highlight1_y}{HTML}{f6a34b}

\definecolor{background0_gray}{HTML}{f2f6f8}
\definecolor{highlight0_gray}{HTML}{dddddd}
\definecolor{highlight1_gray}{HTML}{999999}

\tikzstyle{simplexEdge}=[draw=black, line width=0.1mm]
\tikzstyle{simplexVertex}=[draw=black, thinEdge, fill=white, shape=circle, minimum size=1pt, inner sep=1pt]
\tikzstyle{simplexVertexII}=[line width=0.9mm, shape=circle, minimum size=1.3em, inner sep=2pt]
\tikzstyle{simplexTriangle}=[fill=background0_r, draw=black, line width=0.1mm]

\tikzstyle{simplexVertexBlue}=[draw=black, thinEdge, fill=background0_b, shape=circle, minimum size=1pt, inner sep=1pt]
\tikzstyle{simplexVertexRed}=[draw=black, thinEdge, fill=background0_r, shape=circle, minimum size=1pt, inner sep=1pt]

\tikzstyle{highlightedSimplexVertexR}=[simplexVertexII,draw=highlight1_r,fill=highlight0_r]
\tikzstyle{highlightedSimplexVertexB}=[simplexVertexII,draw=highlight1_b,fill=highlight0_b]
\tikzstyle{highlightedSimplexVertexG}=[simplexVertexII,draw=highlight1_g,fill=highlight0_g]
\tikzstyle{highlightedSimplexVertexY}=[simplexVertexII,draw=highlight1_y,fill=highlight0_y]
\tikzstyle{highlightedSimplexVertexGray}=[simplexVertexII,draw=highlight1_gray,fill=highlight0_gray,veryThinEdge]

\tikzstyle{highlightedSimplexVertexYG}=[simplexVertexII, draw=highlight1_y,dash pattern=on pi*0.75em off pi*0.75em, dash phase=pi*0.75em,postaction={draw=highlight1_g,dash phase=pi*1.5em},  semifill={upper=highlight0_y, lower=highlight0_g, ang=180}]
\tikzstyle{highlightedSimplexVertexYB}=[simplexVertexII, draw=highlight1_y,dash pattern=on pi*0.75em off pi*0.75em, dash phase=pi*0.75em,postaction={draw=highlight1_b,dash phase=pi*1.5em},  semifill={upper=highlight0_y, lower=highlight0_b, ang=180}]
\tikzstyle{highlightedSimplexVertexRG}=[simplexVertexII, draw=highlight1_r,dash pattern=on pi*0.75em off pi*0.75em, dash phase=pi*0.75em,postaction={draw=highlight1_g,dash phase=pi*1.5em},  semifill={upper=highlight0_r, lower=highlight0_g, ang=180}]
\tikzstyle{highlightedSimplexVertexRB}=[simplexVertexII, draw=highlight1_r,dash pattern=on pi*0.75em off pi*0.75em, dash phase=pi*0.75em,postaction={draw=highlight1_b,dash phase=pi*1.5em},  semifill={upper=highlight0_r, lower=highlight0_b, ang=180}]

\tikzstyle{highlightedSimplexEdgeR}=[draw=highlight1_r, thickEdge]
\tikzstyle{highlightedSimplexEdgeB}=[draw=highlight1_b, thickEdge]
\tikzstyle{highlightedSimplexEdgeG}=[draw=highlight1_g, thickEdge]
\tikzstyle{highlightedSimplexEdgeY}=[draw=highlight1_y, thickEdge]

\tikzstyle{highlightedSimplexTriangleR}=[fill=highlight0_r, draw=black, line width=0.1mm]
\tikzstyle{highlightedSimplexTriangleB}=[fill=highlight0_b, draw=black, line width=0.1mm]
\tikzstyle{highlightedSimplexTriangleG}=[fill=highlight0_g, draw=black, line width=0.1mm]
\tikzstyle{highlightedSimplexTriangleY}=[fill=highlight0_y, draw=black, line width=0.1mm]

\tikzstyle{vertWidth} = [line width=0.3mm]
\tikzstyle{vertMinSize} = [minimum size=0.3em]
\tikzstyle{vertInnerSep} = [inner sep=-0.2pt]

\tikzstyle{minimumVertex}=[draw=highlight1_b, vertWidth, fill=highlight0_b, shape=circle, vertMinSize, vertInnerSep]
\tikzstyle{saddleVertex}=[draw=highlight1_g, vertWidth, fill=highlight0_g, shape=circle, vertMinSize, vertInnerSep]
\tikzstyle{maximumVertex}=[draw=highlight1_r, vertWidth, fill=highlight0_r, shape=circle, vertMinSize, vertInnerSep]

\definecolor{msmcolor0}{HTML}{ff80bf}
\definecolor{msmcolor1}{HTML}{ff0000}
\definecolor{msmcolor2}{HTML}{00ff00}
\definecolor{msmcolor3}{HTML}{0000ff}
\definecolor{msmcolor4}{HTML}{ffff00}
\definecolor{msmcolor5}{HTML}{ff00ff}
\definecolor{msmcolor6}{HTML}{00ffff}
\definecolor{msmcolor7}{HTML}{a1a1ff}
\definecolor{msmcolor8}{HTML}{ab8054}

\tikzstyle{PCSimplexR_Active}=[draw=highlight1_r, line width=0.9mm, fill=highlight0_y, shape=circle, minimum size=1.3em, inner sep=2pt]
\tikzstyle{PCSimplexB_Active}=[draw=highlight1_b, line width=0.9mm, fill=highlight0_y, shape=circle, minimum size=1.3em, inner sep=2pt]
\tikzstyle{PCSimplexG_Active}=[draw=highlight1_gray, line width=0.9mm, fill=highlight0_y, shape=circle, minimum size=1.3em, inner sep=2pt]

\tikzstyle{PCSimplexR_Inactive}=[draw=highlight1_r, line width=0.9mm, fill=highlight0_gray, shape=circle, minimum size=1.3em, inner sep=2pt]
\tikzstyle{PCSimplexB_Inactive}=[draw=highlight1_b, line width=0.9mm, fill=highlight0_gray, shape=circle, minimum size=1.3em, inner sep=2pt]
\tikzstyle{PCSimplexG_Inactive}=[draw=highlight1_gray, line width=0.9mm, fill=highlight0_gray, shape=circle, minimum size=1.3em, inner sep=2pt]

\tikzstyle{highlightedSimplexVertexG}=[draw=highlight1_g, line width=0.9mm, fill=highlight0_g, shape=circle, minimum size=1.3em, inner sep=2pt]
\tikzstyle{highlightedSimplexVertexY}=[draw=highlight1_y, line width=0.9mm, fill=highlight0_y, shape=circle, minimum size=1.3em, inner sep=2pt]

\usepackage{amsfonts}
\usepackage{amssymb}

\newcommand\MYhyperrefoptions{bookmarks=true,bookmarksnumbered=true,
pdfpagemode={UseOutlines},plainpages=false,pdfpagelabels=true,
colorlinks=true,linkcolor={black},citecolor={black},urlcolor={black},
pdftitle={Parallel Computation of Piecewise Linear Morse-Smale Segmentations},
pdfsubject={Algorithm/Technique},
pdfauthor={Robin G. C. Maack},
pdfkeywords={Morse-Smale Complex, Watershed transformation, Segmentation, Topology, Visualization}}

\ifCLASSINFOpdf
  \usepackage[\MYhyperrefoptions,pdftex]{hyperref}
\else
  \usepackage[\MYhyperrefoptions,breaklinks=true,dvips]{hyperref}
  \usepackage{breakurl}
\fi

\renewcommand{\figureautorefname}{Fig.}
\renewcommand{\sectionautorefname}{Sec.}

\usepackage{enumitem}

\begin{document}

\title{Parallel Computation of\\Piecewise Linear Morse-Smale Segmentations}

\author{
    Robin G. C. Maack,
    Jonas Lukasczyk, 
    Julien Tierny, 
    Hans Hagen, 
    Ross Maciejewski, 
    and
    Christoph Garth 
    \IEEEcompsocitemizethanks{
    \IEEEcompsocthanksitem R. G. C. Maack, J. Lukasczyk, H. Hagen, and C. Garth\\are with RPTU Kaiserslautern-Landau.\protect\\
    E-mails: \{maack, lukasczyk, hagen, garth\}@rptu.de
    \IEEEcompsocthanksitem J. Tierny is with the CNRS and Sorbonne Universit\'{e}.\protect\\
    Email: julien.tierny@sorbonne-universite.fr
    \IEEEcompsocthanksitem R. Maciejewski is with Arizona State University.\protect\\
    Email: rmacieje@asu.edu}
}

\ifCLASSOPTIONpeerreview
  \markboth{IEEE Transactions on Visualization and Computer Graphics,~Vol.~X, No.~X, Month~202X}%
  {Parallel Computation of Piecewise Linear Morse-Smale Segmentations}
\else 
  \markboth{IEEE Transactions on Visualization and Computer Graphics,~Vol.~X, No.~X, Month~202X}%
  {Maack \MakeLowercase{\textit{et al.}}: Parallel Computation of Piecewise Linear Morse-Smale Segmentations}
\fi

\IEEEtitleabstractindextext{%
\begin{abstract}
This paper presents a well-scaling parallel algorithm for the computation of Morse-Smale (MS) segmentations, including the region separators and region boundaries. The segmentation of the domain into ascending and descending manifolds, solely defined on the vertices, improves the computational time using path compression and fully segments the border region. Region boundaries and region separators are generated using a multi-label marching tetrahedra algorithm. This enables a fast and simple solution to find optimal parameter settings in preliminary exploration steps by generating an MS complex preview. It also poses a rapid option to generate a fast visual representation of the region geometries for immediate utilization. Two experiments demonstrate the performance of our approach with speedups of over an order of magnitude in comparison to two publicly available implementations. The example section shows the similarity to the MS complex, the useability of the approach, and the benefits of this method with respect to the presented datasets. We provide our implementation with the paper.
\end{abstract}

}

\IEEEoverridecommandlockouts
\IEEEpubid{\makebox[\columnwidth]{978-1-5386-5541-2/18/\$31.00~\copyright2018 IEEE \hfill} \hspace{\columnsep}\makebox[\columnwidth]{ }}

\maketitle

\IEEEpubidadjcol

\ifCLASSOPTIONcompsoc
\IEEEraisesectionheading{\section{Introduction}\label{sec:introduction}}
\else
\section{Introduction}
\label{sec:introduction}
\fi

\definecolor{darkgreen}{rgb}{0.0, 0.5, 0.0}
\newcommand{\jonas}[1]{{\color{red}[#1]}}
\newcommand{\julien}[1]{{\color{blue}#1}}
\newcommand{\ross}[1]{{\color{orange}#1}}
\newcommand{\robin}[1]{{\color{darkgreen}#1}}

\renewcommand{\subsectionautorefname}{Section}
\renewcommand{\subsubsectionautorefname}{Section}

\begin{figure}
    \centering
    \newcommand{\sepModeScaleX}{1}
\newcommand{\sepModeScaleY}{1}

\tikzstyle{sepEdge}=[gray,thin]
\tikzstyle{sepNode}=[minimum size=1.5em,inner sep=0em]
\tikzstyle{sR}=[highlightedSimplexVertexR,sepNode]
\tikzstyle{sG}=[highlightedSimplexVertexG,sepNode]
\tikzstyle{sB}=[highlightedSimplexVertexB,sepNode]
\tikzstyle{sO}=[highlightedSimplexVertexY,sepNode]
\tikzstyle{sYG}=[highlightedSimplexVertexYG,sepNode]
\tikzstyle{sYB}=[highlightedSimplexVertexYB,sepNode]
\tikzstyle{sRG}=[highlightedSimplexVertexRG,sepNode]
\tikzstyle{sRB}=[highlightedSimplexVertexRB,sepNode]
\tikzstyle{sX}=[highlightedSimplexVertexGray,sepNode]


\newcommand{\sepModeNodes}[2]{
    \foreach \a/\b/\c/\d/\e/\f/\g/\h/\i/\j/\k/\l/\m/\n/\o/\p in {#1}{
        \foreach \ac/\bc/\cc/\dc/\ec/\fc/\gc/\hc/\ic/\jc/\kc/\lc/\mc/\nc/\oc/\pc in {#2}{
            
            \node[\ac] at (0.0,0.0) {\a};
            \node[\bc] at (1.0,0.0) {\b};
            \node[\cc] at (2.0,0.0) {\c};
            \node[\dc] at (3.0,0.0) {\d};
            
            \node[\ec] at (0.0,1.0) {\e};
            \node[\fc] at (1.0,1.0) {\f};
            \node[\gc] at (2.0,1.0) {\g};
            \node[\hc] at (3.0,1.0) {\h};
            
            \node[\ic] at (0.0,2.0) {\i};
            \node[\jc] at (1.0,2.0) {\j};
            \node[\kc] at (2.0,2.0) {\k};
            \node[\lc] at (3.0,2.0) {\l};
            
            \node[\mc] at (0.0,3.0) {\m};
            \node[\nc] at (1.0,3.0) {\n};
            \node[\oc] at (2.0,3.0) {\o};
            \node[\pc] at (3.0,3.0) {\p};
        }
    }
}

\newcommand{\sepModeGrid}{
    \foreach \x in {0,...,3} {
        \draw[sepEdge] (\x,0)--(\x,3);
        \draw[sepEdge] (0,\x)--(3,\x);
    }
    \draw[sepEdge] (0,0)--(3,3);
    \draw[sepEdge] (0,1)--(2,3);
    \draw[sepEdge] (0,2)--(1,3);
    \draw[sepEdge] (1,0)--(3,2);
    \draw[sepEdge] (2,0)--(3,1);
}

\newcommand{\drawSepNodes}[1]{
    \foreach \x/\y in {#1}{
        \node[simplexVertex] at (\x,\y) {};
    }
}

\newcommand{\sepCapY}{-0.65}

\begin{tikzpicture}[xscale=\sepModeScaleX,yscale=\sepModeScaleY]
    \begin{scope}[xshift=0em,yshift=0em]
        \sepModeGrid
        \sepModeNodes
            {14/12/2 /0 /13/11/3 /4 /5 /6 /7 /8 /1 /9 /10/15}%
            {sB/sX/sX/sR/sX/sX/sX/sX/sX/sX/sX/sX/sO/sX/sX/sG}
        \node at (1.5,\sepCapY) {a) Input Scalar Field};
    \end{scope}
    
    \begin{scope}[xshift=-6.5em,yshift=-14em]
        \sepModeGrid
        \sepModeNodes
            {14/12/2 /0 /13/11/3 /4 /5 /6 /7 /8 /1 /9 /10/15}%
            {sB/sB/sB/sG/sB/sB/sB/sG/sB/sB/sG/sG/sG/sG/sG/sG}
            
        \draw[thick] (0,2.5)--(1.5,2.5)--(1.5,1.5)--(2.5,1.5)--(2.5,0);
        \drawSepNodes{0.0/2.5, 0.5/2.5, 1.0/2.5, 1.5/2.5, 1.5/2.0, 1.5/1.5, 2.0/1.5, 2.5/1.5, 2.5/1.0, 2.5/0.5, 2.5/0.0}
        
        \node at (0,\sepCapY-0.2) {b)};
        \node at (1.7,\sepCapY) {Descending Seg.};
        \node at (1.7,\sepCapY-0.4) {+ Region Separators};
    \end{scope}
    
    \begin{scope}[xshift=6.5em,yshift=-14em]

        \draw[fill=highlight0_gray] (0,0)--(3,3)--(2,3)--(0,1)--cycle;
        \sepModeGrid
        \draw[line width=0.2em] (0,0)--(3,3);
        \draw[line width=0.2em] (0,1)--(2,3);
        
        \sepModeNodes
            {14/12/2 /0 /13/11/3 /4 /5 /6 /7 /8 /1 /9 /10/15}%
            {sR/sR/sR/sR/sO/sR/sR/sR/sO/sO/sR/sR/sO/sO/sO/sR}
            
        
        \node at (0,\sepCapY-0.2) {c)};
        \node at (1.7,\sepCapY) {Ascending Seg.};
        \node at (1.7,\sepCapY-0.4) {+ Region Boundaries};
    \end{scope}
    
    \begin{scope}[xshift=-6.5em,yshift=-29em]
        \sepModeGrid
        \sepModeNodes
            {14/12/2 /0 /13/11/3 /4 /5 /6 /7 /8 /1 /9 /10/15}%
            {sRB/sRB/sRB/sRG/sYB/sRB/sRB/sRG/sYB/sYB/sRG/sRG/sYG/sYG/sYG/sRG}
            
        \draw[thick] (0,2.5)--(1.5,2.5)--(1.66,2.33)--(1.33,1.66)--(1.5,1.5)--(2.5,1.5)--(2.5,0.0);
        \draw[thick] (0,0.5)--(1,1.5)--(1.33,1.66);
        \draw[thick] (1.66,2.33)--(2.0,2.5)--(2.5,3.0);
        \drawSepNodes{0/2.5,1.5/2.5,1.66/2.33,1.33/1.66,1.5/1.5,2.5/1.5,2.5/0.0}
        \drawSepNodes{0/0.5,0.5/1.0,1.0/1.5,0.5/2.5,1.0/2.5,2.0/1.5,2.5/1.0,2.5/0.5,2.0/2.5,2.5/3.0,1.5/2.0}
        
        \node at (0,\sepCapY-0.2) {d)};
        \node at (1.7,\sepCapY) {Morse-Smale Seg.};
        \node at (1.7,\sepCapY-0.4) {+ Region Separators};
    \end{scope}
    
    \begin{scope}[xshift=6.5em,yshift=-29em]
        \draw[fill=highlight0_gray] (0,2)--(0,3)--(3,3)--(2,2)--(3,2)--(3,0)--(2,0)--(2,1)--(1,1)--(0,0)--(0,1)--(1,2)--cycle;
        \sepModeGrid
            
        \draw[line width=0.2em] (0,3)--(2,3);
        \draw[line width=0.2em] (0,2)--(1,2)--(0,1);
        \draw[line width=0.2em] (0,0)--(1,1)--(2,1)--(2,0);
        \draw[line width=0.2em] (3,0)--(3,2)--(2,2)--(3,3);
        
        \sepModeNodes
            {14/12/2 /0 /13/11/3 /4 /5 /6 /7 /8 /1 /9 /10/15}%
            {sRB/sRB/sRB/sRG/sYB/sRB/sRB/sRG/sYB/sYB/sRG/sRG/sYG/sYG/sYG/sRG}
        
        \node at (0,\sepCapY-0.2) {e)};
        \node at (1.7,\sepCapY) {Morse-Smale Seg.};
        \node at (1.7,\sepCapY-0.4) {+ Region Boundaries};
    \end{scope}   
\end{tikzpicture}
\vspace{10pt}
    \vspace*{-3em}
    \caption{A $4\times4$ example triangulation showing the input field (a) and available segmentations and region-separating geometries (b-e). Each color represents the influence area of a specific extremum (the areas of minima 0 and 1 are shown in red and yellow, while the area of maxima 14 and 15 are shown in blue and green). The descending segmentation (b) represents the influence area of maxima and the ascending segmentation (c) the ones of all minima. In the case of Morse-Smale segmentations (d-e) the nodes show the influence of minima-maxima combinations, leading to nodes being colored by the minima color at the bottom and the maximum color at the top. Subfigures (b) and (d) show region separators (thin black lines), while (c) and (e) show region boundaries (thick black lines).
    \label{fig:wallmodes}
    }
\end{figure}
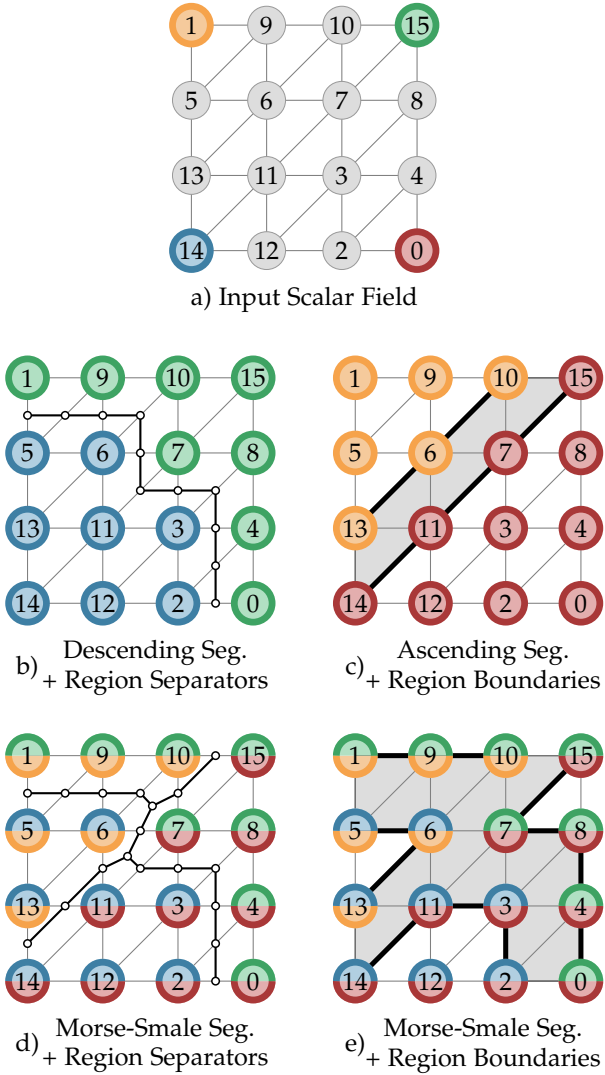

\IEEEPARstart{T}{opological Data Analysis}~(TDA) provides a family of effective feature characterizations, including the well-studied Morse-Smale~(MS) complex. 
The MS complex is a central tool in TDA for feature-driven data analysis and visualization, as it segments the domain of a scalar field into regions with equivalent gradient flow behavior (see \autoref{fig:wallmodes} and \autoref{sec:Preliminaries}).
This rather abstract feature characterization based on gradient flow has been applied successfully in several domains, including
chemistry~\cite{olejniczak2020topological,bhatia2018topoms},
material science~\cite{venkat2021towards,homberg2014definition}, 
physics~\cite{laney2006understanding,gyulassy2015interstitial},
and
cosmology~\cite{sousbie2011persistent}.

However, due to the high computational complexity of MS complex generation, it often becomes a time-consuming bottleneck. Especially in the case of interactive analysis and visualization, users expect to quickly retrieve a visual output, where long wait times can interrupt their workflow.
Furthermore, many applications do not require the computation of the full MS complex, but only two of its central features:
1)~the MS segmentation of the domain that assigns extrema to each vertex by following the gradient along the steepest descend and ascend;
and
2)~the interfaces between different regions in the segmentation.  

In this paper we describe a scalable implementation of these two tasks with high parallel efficiency, further referred to as the piecewise linear Morse-Smale segmentation~(PLMSS) algorithm~(\autoref{sec:Methods}).
As the name suggests, the input of PLMSS is a scalar field defined on the vertices of a piecewise linear domain, i.e., a simplicial complex.
PLMSS utilizes path compression to derive the MS segmentation of the domain and a multi-label marching tetrahedra procedure to derive the interfaces between different regions of the MS segmentation.
We chose these two underlying algorithms since they are known to be embarrassingly parallelizable, and therefore expected to scale well.
In \autoref{sec:Results} we demonstrate the benefit of using PLMSS for effective data analysis and visualization on four datasets.

We compare PLMSS against the corresponding subprocedures of two state-of-the-art Morse-Smale complex software libraries, i.e., the implementations available in the Topology ToolKit~(TTK)~\cite{ttk19} and MSCEER~\cite{msceer}.
Specifically, we performed two strong scaling studies~(\autoref{sec:performance}) that show that MSCEER outperforms TTK, but PLMSS is still up to an order of magnitude faster than MSCEER, while also providing superior parallel efficiency.
However, the qualitative and quantitative comparison between the region separators computed by PLMSS and their counterpart of MS complex 2-cells is more challenging.
Although they both separate regions with different gradient flow behavior, they are defined and computed differently~(\autoref{sec:Preliminaries}).
Yet, for many applications---including the scenarios presented in \autoref{sec:Results}---the computation of region separators is sufficient without the need for expensive discrete gradient field and dual mesh computations.\vspace*{0.4em}

\noindent In short, the contributions of our work are:
\begin{itemize}[itemindent=0em]
    \item A scalable algorithm with high parallel efficiency for the computation of piecewise linear Morse-Smale segmentations~(PLMSS);
    \item A detailed performance benchmark that compares PLMSS with the corresponding subprocedures of two state-of-the-art Morse-Smale complex software libraries (TTK and MSCEER); and
    \item The integration of PLMSS in TTK to facilitate future benchmarks and reproducibility.
\end{itemize}

\section{Related Work}
\label{sec:RelatedWork}
The MS complex subdivides a given scalar field into regions of uniform gradient flow behavior, segmenting the domain such that each point in the same MS manifold will flow towards the same critical point pair considering forward and backward integration. Two approaches to computing the MS complex arose from Morse theory~\cite{milnor2016morse}, where either a discrete gradient vector field is defined on the whole domain~\cite{forman1998morse}, or piecewise linear Morse theory is used to define a segmentation~\cite{banchoff1967critical, banchoff1970critical}. For an in-depth comparison of both approaches, we refer to Lewiner's work~\cite{lewiner2013critical}.

Algorithms based on piecewise linear Morse theory are divided into boundary-based and region-growing algorithms. Boundary-based algorithms trace lines of steepest descent/ascent seeded at the saddles, such that every vertex on a line of steepest ascent/descent belongs to the same region. Region-growing algorithms grow sets of top-level cells, e.g. cubes or tetrahedra in 3D, located at the minima and maxima of a scalar function, iteratively enlarging regions. One representative of boundary-based algorithms is Edelsbrunner et al.~\cite{edelsbrunner2001hierarchical} that first introduced the MS complex for piecewise linear 2-manifolds, recording paths of steepest ascent and descent. They also introduced the notion of the quasi MS complex that was extended to 3-manifolds~\cite{edelsbrunner2003morse} and later improved in geometric accuracy by Bremer et al.~\cite{bremer2004topological}. Concerning region-growing algorithms, Danovaro et al.~\cite{danovaro2003topological} started growing regions by using triangles incident on maxima at vertices, adding edge incident triangles iteratively. They extended this approach by appending additional seeding points at initialization, while also enabling to process higher-dimensional scalar fields~\cite{danovaro2003morphology}. Gyulassy et al.~\cite{gyulassy2007efficient} implemented a region-growing algorithm that labels the vertices to extract 3-, 2-, and 1-cells, also extendable from 3D to higher dimensional scalar fields.

Discrete Morse theory was developed by Forman~\cite{forman2002user}, applying Morse theory to any type of simplicial cell complex. Many algorithms build upon the discrete gradient vector field to effectively compute the MS complex~\cite{gunther2012efficient, robins2011theory, king2005generating}. To efficiently compute the MS complex for large datasets, several parallel and distributed memory implementations were introduced. Gyulassy et al.~\cite{gyulassy2008practical} first proposed splitting a dataset into subsets called parcels, extracting the MS complex from each parcel to later merge them in a cancellation-based step. This allowed for the computation of datasets that do not fit into memory and gave rise to distributed memory approaches~\cite{gyulassy2012parallel, peterka2011scalable}. Further parallel optimizations were achieved by merging gradient paths, enabling the computation of the gradient assignment and extrema traversals on the GPU~\cite{shivashankar2012parallel, shivashankar2012parallel3D}. Subhash et al.~\cite{subhash2020gpu} then accomplished computing all steps of the MS complex computation on the GPU. Even though some algorithms improved the steepest descent line tracing~\cite{bremer2004topological, gyulassy2007efficient} by allowing the traversal to use edges and triangles, still all presented algorithms often produce incorrect connectivity and inaccurate geometry due to the refinement of the underlying discrete domain~\cite{heine2016survey}. Here, Gyulassy et al.~\cite{gyulassy2012computing} implemented a probabilistic algorithm to extract the correct geometry and connectivity. Morse-Smale complexes were already used in many applications such as material science~\cite{gyulassy2007topologically}, chemistry~\cite{gunther2014characterizing}, and medicine~\cite{gyulassy2014conforming}, allowing for fast and consistent analysis of the data. Cancellation-based simplification~\cite{lukasczyk2020localized, fellegara2014efficient, weinkauf2010topology} is often used in applications, where pairs of critical points are removed to simplify the MS complex and eliminate noise in the dataset. It counteracts over-segmentation of the domain and enables the extraction of persistent features.

The Watershed transform, originally defined by Beucher and Lantu\'{e}joul~\cite{beucher1979use}, is another approach to Morse theory, segmenting the domain, usually gray-scale images, into catchment basins that represent the zone of influence of minima and watershed lines, separating catchment basins from each other. Beucher~\cite{beucher1982watersheds} described catchment basins as areas where each drop of water ends up at the same minimum when flowing down the surface. In contrast to Morse theory, minima don't have to be distinct but can rather consist of multiple vertices of the same function value. Those definitions were further improved to be rigorous~\cite{meyer1993integrals, najman1993definition} and extended to the discrete case~\cite{meyer1994topographic, vincent1991watersheds}. De Floriani~\cite{de2015morse} distinguishes three types of watershed algorithms that are either based on topographic distance, simulated immersion, or rainfalling simulation. Algorithms based on topographic distance compute shortest paths to find corresponding catchment basins~\cite{meyer1990morphological, meyer1994topographic}. Gabrielyan et al.~\cite{gabrielyan2022parallel} and Yeghiazaryan and Voiculescu~\cite{yeghiazaryan2018path} provide GPU implementations using an approach similar to path compression, speeding up the shortest path computation. Simulated immersion approaches seed catchment basins at minima, extending them by processing vertices in increasing function value order\cite{vincent1991watersheds, soille2004morphological}. Rainfalling simulation algorithms use an inverted logic, finding and labeling minima first, then decreasing in function value for each vertex by steepest descent until a labeled vertex is found~\cite{mangan1999partitioning, stoev2000extracting}.

Marching tetrahedra~\cite{marchingTetrahedra} is an algorithm to extract boundary surfaces from a tetrahedralization separating differently labeled vertices from each other. It is a generalization of the marching cubes algorithm~\cite{marchingCubes,marchingCubesAmbiguity} that initially allowed the creation of iso-surfaces at a selected isovalue by subdividing voxels such that areas with values above and below the isovalue were separated. Yet, in contrast to the original marching cubes algorithm, it allows multiple labels to be present at each tetrahedron and always extracts a distinct triangulation at each tetrahedron, eliminating ambiguous cases. The effect of various simplicial subdivisions on the quality of the resulting surfaces has been studied by Carr et al.\cite{carr2006artifacts}. Also, various applications use this approach for its simplicity and performance~\cite{chouchaneVersatile2019, MULLER2014205,weinstein2000,zhang2006integrating}.

\section{Preliminaries}
\label{sec:Preliminaries}

\newcommand{\discreteVectorField}{\mathcal{V}}
\newcommand{\numberProcesses}{n_p}
\newcommand{\ghostLayer}{\mathcal{G}}
\newcommand{\domain}{\mathcal{M}}
\newcommand{\range}{\mathbb{R}}
\newcommand{\sublevelset}[1]{#1^{-1}_{-\infty}}
\newcommand{\superlevelset}[1]{#1^{-1}_{+\infty}}
\newcommand{\Star}{St}
\newcommand{\Link}{Lk}
\newcommand{\simplex}{\sigma}
\newcommand{\face}{\tau}
\newcommand{\lowerlink}{\Link^{-}}
\newcommand{\upperlink}{\Link^{+}}
\newcommand{\Index}{\mathcal{I}}
\newcommand{\offset}{o}
\newcommand{\Natural}{\mathbb{N}}
\newcommand{\criticalSet}{\mathcal{C}}
\newcommand{\diagram}{\mathcal{D}}
\newcommand{\wasserstein}[1]{W^{\diagram}_#1}
\newcommand{\projection}{\Delta}
\newcommand{\hierarchy}{\mathcal{H}}
\newcommand{\decimation}{D}
\newcommand{\xDimD}{L_x^\decimation}
\newcommand{\yDimD}{L_y^\decimation}
\newcommand{\zDimD}{L_z^\decimation}
\newcommand{\xDim}{L_x}
\newcommand{\yDim}{L_y}
\newcommand{\zDim}{L_z}
\newcommand{\Grid}{\mathcal{G}}
\newcommand{\GridD}{\mathcal{G}^\decimation}
\newcommand{\x}{\phantom{x}}
\newcommand{\Mod}{\;\mathrm{mod}\;}
\newcommand{\NN}{\mathbb{N}}
\newcommand{\forwardIntegralLine}{\mathcal{L}^+}
\newcommand{\backwardIntegralLine}{\mathcal{L}^-}
\newcommand{\triangulationOp}{\phi}
\newcommand{\decimationOp}{\Pi}
\newcommand{\isovalue}{w}
\newcommand{\persistence}{\mathcal{P}}
\newcommand{\pointMetric}{d}
\newcommand{\diagramSet}{\mathcal{S}_\mathcal{D}}
\newcommand{\diagramSpace}{\mathbb{D}}
\newcommand{\jointree}{\mathcal{T}^-}
\newcommand{\splittree}{\mathcal{T}^+}
\newcommand{\mergetree}{\mathcal{T}}
\newcommand{\mergetreeSet}{\mathcal{S}_\mathcal{T}}
\newcommand{\branchset}{\mathcal{S}_\mathcal{B}}
\newcommand{\branchspace}{\mathbb{B}}
\newcommand{\mergetreeSpace}{\mathbb{T}}
\newcommand{\editdistance}{D_E}
\newcommand{\wassersteinTree}{W^{\mergetree}_2}
\newcommand{\distanceSequence}{d_S}
\newcommand{\branchtree}{\mathcal{B}}
\newcommand{\branchtreeSet}{\mathcal{S}_\mathcal{B}}
\newcommand{\branchtreeSpace}{\mathbb{B}}
\newcommand{\forest}{\mathcal{F}}
\newcommand{\sequenceSpace}{\mathbb{S}}
\newcommand{\forestMatrix}{\mathbb{F}}
\newcommand{\treeMatrix}{\mathbb{T}}
\newcommand{\normalizedLocation}{\mathcal{N}}
\newcommand{\normalizedWasserstein}{W^{\normalizedLocation}_2}
\newcommand{\geodesictree}{\mathcal{G}}
\newcommand{\dummyVector}{\mathcal{V}}
\newcommand{\geodesictreeVec}{g}
\newcommand{\geodesicAxis}{\mathcal{A}}
\newcommand{\directionVector}{\mathcal{V}}
\newcommand{\geodesicdiagram}{\mathcal{G}^{\diagram}}
\newcommand{\reconstructionError}{E_{L_2}}
\newcommand{\pcaBasis}{B_{\mathbb{R}^d}}
\newcommand{\mtPgaBasis}{B_{\branchtreeSpace}}
\newcommand{\mtPgaError}{E_{\wassersteinTree}}
\newcommand{\frechetEnergy}{E_F}
\newcommand{\geodesicExtremity}{\mathcal{E}}
\newcommand{\vectorNotation}[1]{\protect\vv{#1}}
\newcommand{\axisNotation}[1]{\protect\overleftrightarrow{#1}}
\newcommand{\individualEnergy}{E}
\newcommand{\ensembleSize}{N}
\newcommand{\numberBranchinBarycenter}{N_1}
\newcommand{\numberGeodesicSamples}{N_2}
\newcommand{\planarGridX}{N_x}
\newcommand{\planarGridY}{N_y}
\newcommand{\regularGrid}{G}
\newcommand{\distanceMatrix}{\mathbb{D}}
\newcommand{\maxDimensions}{{d_{max}}}
\newcommand{\projectionOperator}{\mathcal{P}}
\newcommand{\reconstructed}[1]{\widehat{#1}}
\newcommand{\gt}{>}
\newcommand{\lt}{<}
\newcommand{\branch}{b}

\renewcommand{\figureautorefname}{Fig.}
\renewcommand{\sectionautorefname}{Sec.}
\renewcommand{\subsectionautorefname}{Sec.}
\renewcommand{\subsubsectionautorefname}{Sec.}
\renewcommand{\equationautorefname}{Eq.}
\renewcommand{\tableautorefname}{Tab.}
\newcommand{\algorithmautorefname}{Alg.}
\newcommand{\lineautorefname}{Alg.}

\newcommand{\eqSpace}{-1.75ex}

\newcommand{\mycaption}[1]{
\caption{#1}
}

This section describes the formal setting of our work. It contains definitions adapted from the Topology ToolKit (TTK)~\cite{ttk17, ttk19}. We refer the reader to textbooks~\cite{edelsbrunner09, zomorodianBook} for comprehensive introductions to computational topology.

\subsection{Input Data}
The input is a piecewise linear (PL) scalar field $f :
\domain
\rightarrow \mathbb{R}$ defined on a $d$-dimensional simplicial
complex, with $d \leqslant 3$ in our applications.
The \emph{star} $\Star(\simplex)$ of a simplex $\simplex$ is the set of
simplices of $\domain$ which contain $\sigma$ as a face. The \emph{link}
$\Link(\simplex)$ is the set of faces of the simplices of $\Star(\simplex)$
which do not intersect $\simplex$.
The input field $f$ is  provided on the vertices of $\domain$
and interpolated on the simplices of higher dimensions. $f$ is assumed to
be injective, which is achieved in practice by substituting the $f$ value
of a vertex by its position in the non-ambiguous, global vertex order
(by increasing $f$ values).

\subsection{Critical Points}
The sub-level set
$\sublevelset{f}(w)$
of an isovalue $w \in \mathbb{R}$ is
defined as $\sublevelset{f}(w) = \{p \in \mathcal{M} ~|~ f(p) < w\}$. 
As $w$ continuously increases, the topology of $\sublevelset{f}(w)$ changes at
specific vertices of $\domain$, called the \emph{critical points} of $f$.
Let $\Link^-(v)$ be the \emph{lower link} of the vertex $v$:
$\Link^-(v) = \{\simplex \in \Link(v) ~|~ \forall u \in
\simplex : f(u) < f(v)\}$.
The \emph{upper link} of $v$ is defined symmetrically:
$\Link^+(v) = \{\simplex \in \Link(v) ~|~ \forall u \in
\simplex : f(u) > f(v)\}$.
A vertex $v$ is \emph{regular} if and only if both $\Link^-(v)$ and
$\Link^+(v)$ are simply connected. Otherwise, $v$ is a \emph{critical vertex} of
$f$~\cite{banchoff1970critical}.
A critical vertex $v$ can be classified by its \emph{index} $\Index(v)$, which
is
$0$ for minima, $1$ for $1$-saddles, $(d-1)$ for $(d-1)$-saddles and $d$
for maxima. Vertices for which the number of connected components of
$\Link^-(v)$ or $\Link^+(v)$ are greater than $2$ are called \emph{degenerate
saddles}.

\subsection{Integral Lines}
\label{sec_integralLines}
\emph{Integral lines} are piecewise linear curves on 
$\domain$ which
locally describe the gradient of $f$.
They can be used
to capture and visualize
adjacency relations between critical points.
Given a vertex $v$, its \emph{forward} integral line, noted
$\forwardIntegralLine(v)$, is a path along the edges of $\domain$, initiated in
$v$, such that each edge of $\forwardIntegralLine(v)$ connects a vertex $v'$ to
its highest neighbor $v''$. Then forward integrals are guaranteed to terminate
in local maxima of $f$.
A \emph{backward} integral line, noted
$\backwardIntegralLine(v)$, is defined symmetrically (i.e. integrating
downwards towards minima).

Moreover, we define a \emph{forward extremal integral line} as a forward integral line started at a connected component of upper link $\Link^+(s)$ of a saddle $s$. Backward extremal integral lines are defined symmetrically. 
We say that a saddle $s$ is a \emph{forward separating saddle} if there exist at least two forward extremal integral lines starting at $s$ which terminate in distinct local maxima. Backward-separating saddles are defined symmetrically.
In practice, extremal integral lines help capture adjacency relations between critical points.

\subsection{Morse-Smale Segmentation}
\label{sec_segmentation}
In this section, we formalize the notion of Morse-Smale segmentation computed by our approach.

For a given vertex $v$, let $m$ and $M$ be its \emph{integration extremities}: $m$ is the local minimum reached by the backward integral line started in $v$, while $M$ is the local maximum reached by the forward integral line started in $v$. We now introduce an equivalence relation $v_1 \sim v_2$ 
between two vertices $v_1$ and $v_2$, which holds if their integration extremities are identical. The \emph{Morse-Smale (MS) segmentation} is then a decomposition of the set of vertices of $\domain$ into maximal subsets $\domain_i$, called \emph{MS regions}, such that for all pairs of vertices $(v_1, v_2) \in \domain_i$, we have $v_1 \sim v_2$.

Let $\tau$ be a $d'$-simplex of $\domain$ (with $0 \leq d' < d$), which only contains vertices belonging to a single MS region $\domain_i$. If the link $\Link(\tau)$
includes vertices which do \emph{not} belong to $\domain_i$, we say that $\tau$ is a boundary simplex for $\domain_i$. Then the \emph{region boundary} of $\domain_i$ is the simplicial complex formed by the union of all the boundary simplices of $\domain_i$ (and their faces). Each \emph{region boundary} separates $\domain_i$ from the remaining dataset. The \emph{region separators} separate all regions $\domain_i \in \domain$ from each other. To create them, every $d$-simplex of $\domain$ that contains vertices belonging to at least two distinct MS regions spawns $(d-1)$-simplices inside its convex hull, as depicted in \autoref{fig:TriangleCases} and \autoref{fig:tetrahedraTriangulation}.

\subsection{Discrete Morse Theory}
We now conclude this section of preliminaries with notions (adapted from~\cite{guillou_tech22}) of discrete Morse theory~\cite{forman1998morse}, or DMT for short, as it has become a central component in modern implementations of the notion of Morse-Smale complex.
We discuss the key differences between the Morse-Smale complex and the structures extracted by our approach (formalized in Secs. \ref{sec_integralLines} and \ref{sec_segmentation}).

A \emph{discrete vector} is a pair formed by a simplex $\simplex_i \in
\domain$ (of dimension $i$) and one of its co-facets $\simplex_{i+1}$ (i.e. one
of its co-faces of dimension $i+1$), noted $\{\simplex_i < \simplex_{i+1}\}$.
$\simplex_{i+1}$ is usually referred to as the \emph{head} of the vector, while
$\simplex_{i}$ is its \emph{tail}. Examples of discrete vectors include a pair
between a vertex and one of its incident edges or a pair between an edge and a
triangle containing it. A \emph{discrete vector field} on $\domain$ is then
defined as a collection $\discreteVectorField$ of pairs $\{\simplex_i <
\simplex_{i+1}\}$, such that each simplex of $\domain$ is involved in at most
one pair. A simplex $\simplex_i$ which is involved in no discrete vector
$\discreteVectorField$ is called a \emph{critical simplex}.

A \emph{v-path} is a sequence of discrete vectors
$\big\{\{\simplex^0_i < \simplex^0_{i+1}\}, \dots,
\{\simplex^k_i <
\simplex^k_{i+1}\}\big\}$, such that \emph{(i)} $\sigma^j_i \neq
\sigma^{j+1}_i$ (i.e. the tails of two consecutive vectors are distinct) and
\emph{(ii)} $\sigma^{j+1}_i < \sigma^{j}_{i+1}$ (i.e. the tail of a vector in
the sequence is a face of the head of the previous vector), for any $0 < j <
k$.
A \emph{v-path} can be interpreted as the discrete analog to the notion of PL integral line introduced in \autoref{sec_integralLines}.
We say that a v-path \emph{terminates} at a critical simplex $\simplex_i$ if $\simplex_i$ is a face of the head of its last vector $\{\simplex^k_i < \simplex^k_{i+1}\}$. Symmetrically, we say that a v-path \emph{starts} at a critical simplex $\simplex_{i+1}$ if $\simplex_{i+1}$ is a co-facet of the tail of its first vector $\{\simplex^0_i < \simplex^0_{i+1}\}$. Then, the collection of all the v-paths terminating in a given critical simplex $\simplex_i$ is called the \emph{discrete stable set} of $\simplex_i$ and is noted $\domain(\simplex_i)$. Symmetrically, the collection of all the v-path starting at a given critical simplex $\simplex_i$ is called the \emph{discrete unstable set} of $\simplex_i$ and is noted $\domain'(\simplex_i)$.

A \emph{discrete gradient field} is then a discrete vector field such that
all its possible \emph{v-paths} are loop-free. Several algorithms have been
proposed to compute such a discrete gradient field from an
input PL scalar field (see~\cite{robins2011theory} for instance).
The \emph{discrete Morse complex} is then defined as the complex formed by the discrete unstable sets of all the critical simplices. It is a cell complex made of $d'$-dimensional cells (with $d' \in \{0, 1, \dots, d\}$), such that each $d'$-dimensional cell is the discrete unstable set of a critical $d'$-simplex.
The \emph{opposite discrete Morse complex} is defined symmetrically, i.e. it is the cell complex formed by the discrete stable sets of all the critical simplices.
Finally, the \emph{discrete Morse-Smale complex} is defined as the complex formed by the intersections of the cells of the  {discrete Morse complex} and the {opposite discrete Morse complex}. 

Several conceptual differences exist between the Morse-Smale complex and the Morse-Smale (MS) segmentations considered in our work. First, as their name suggests, MS segmentations only provide vertex-based decompositions of the input domain,  not a cell complex that exhaustively and precisely captures all possible adjacency relations between integral lines (formally v-paths). Thus, MS segmentations target a subset of the applications enabled by the Morse-Smale complex (specifically, involving data segmentation). While the separatrices of the MS regions (\autoref{sec_segmentation}) resemble the $2$-dimensional cells of the Morse-Smale complex, they only correspond to the unstable sets of \emph{separating} saddles (\autoref{sec_integralLines}), which constitutes a subset of all the saddles (i.e. saddles where isosurfaces change their genus are not considered). 
Finally, note that in DMT, local maxima (critical $d$-simplices) cannot strictly occur on the boundary of $\domain$, which only includes $d'$-simplices (with $d' < d$). 

\section{Method}
\label{sec:Methods}
In this section, the algorithms for the computation of the PLMSS are described in detail. First, necessary preprocessing steps and data structures are presented. Then the ascending and descending segmentation of the domain is described, followed by the computation of the MS segmentation. 

\subsection{Preprocessing}
\label{sec:preprocessing}

To prevent ambiguity during the computation of integral paths, we apply a variant of \emph{Simulation of Simplicity}~\cite{edelsbrunner1990simulation} on the input scalar field $f$.
We first sort all vertices of the domain according to their scalar value, where we resolve ties based on the indices of the compared vertices.
Then, we derive the so-called order field $\bar f$ that records for each vertex its index in this sorted array.
Note, that each critical point of $f$ is also a critical point of $\bar f$, but $\bar f$ might exhibit additional critical points that result from the disambiguation.
These spurious critical points, however, can be removed via topological simplification, which we apply in order to remove non-persistent critical points from the scalar field.
For a detailed discussion on topological simplification and its implementation in TTK, we refer the reader to the work of Lukasczyk et al.~\cite{lukasczyk2020localized}.

The advantage of processing an order field over the original input scalar field is that $\bar f$ is injective, i.e., every vertex has a distinct largest and smallest neighbor in the order field.
It is only possible that a vertex has no neighbor with a larger or smaller order value, in which case the vertex is a maximum or minimum, respectively. Hence, there is always a distinct direction of steepest ascent and descent, which is essential for the computation of the ascending and descending manifolds, described next.

\subsection{Segmentation and Extrema Retrieval}
\label{sec:segmentation}

The segmentation of the domain is a two-step process. In the first step, the ascending and descending segmentations are created; representing areas of influence of minima and maxima, respectively. These segmentations are intersected to create the MS segmentation, representing the areas of influence of minimum-maximum pairs.

\subsubsection{Ascending and Descending Segmentation}
\label{sec:method_segmentation}

MS segmentations subdivide a domain into areas of similar flow behavior, meaning that forward and backward integration for any vertex in the same region leads to the same extremum pair. This means that each MS subset of the domain corresponds to all steepest descent/ascent paths that terminate in the same pair of extrema. To achieve this, first, every vertex has to be assigned to its minimum and maximum. Therefore, the ascending ($asc$) and descending ($dsc$) segmentations of the domain are computed. As the process is the same for both directions, without loss of generality, it will be described for the descending segmentation.

\begin{figure}[!htb]
    \centering
    \newcommand{\pathCompressionSpace}{1cm}
\newcommand{\pathCompressionNodes}[8]{
    \node[right] (Label) at (-2.3,0)  {#1};
    \node[#2] (A) at (0.0,0.0) {0};
    \node[#3] (B) at (1.0,0.0) {1};
    \node[#4] (C) at (2.0,0.0) {2};
    \node[#5] (D) at (3.0,0.0) {3};
    \node[#6] (E) at (4.0,0.0) {4};
    \node[#7] (F) at (5.0,0.0) {5};
    \node[#8] (G) at (6.0,0.0) {6};
}

\begin{tikzpicture}
\pathCompressionNodes{Gradient:}{highlightedSimplexVertexGray}{highlightedSimplexVertexGray}{highlightedSimplexVertexGray}{highlightedSimplexVertexGray}{highlightedSimplexVertexGray}{highlightedSimplexVertexGray}{highlightedSimplexVertexR}

\draw[->] (A) to (B);
\draw[->] (B) to (C);
\draw[->] (C) to (D);
\draw[->] (D) to (E);
\draw[->] (E) to (F);
\draw[->] (F) to (G);

\node[] (Space) at (0.0,-0.5) {};
\end{tikzpicture}
\begin{tikzpicture}
\pathCompressionNodes{Iteration 1:}{highlightedSimplexVertexGray}{highlightedSimplexVertexGray}{highlightedSimplexVertexGray}{highlightedSimplexVertexGray}{PCSimplexR_Inactive}{PCSimplexR_Inactive}{highlightedSimplexVertexR}

\draw[->] (A) to [out=-40,in=-140] (C);
\draw[->] (B) to [out=40,in=140] (D);
\draw[->] (C) to [out=-40,in=-140] (E);
\draw[->] (D) to [out=40,in=140] (F);
\draw[->] (E) to [out=-40,in=-140] (G);
\draw[->] (F) to (G);

\node[] (Space) at (0.0,-0.5) {};
\end{tikzpicture}
\begin{tikzpicture}
\pathCompressionNodes{Iteration 2:}{highlightedSimplexVertexGray}{highlightedSimplexVertexGray}{PCSimplexR_Inactive}{PCSimplexR_Inactive}{PCSimplexR_Inactive}{PCSimplexR_Inactive}{highlightedSimplexVertexR}

\draw[->] (B) to [out=30,in=150] (F);
\draw[line width=0.3em,white] (A) to [out=30,in=150] (E);
\draw[->] (A) to [out=30,in=150] (E);
\draw[->] (C) to [out=-30,in=-150] (G);
\draw[->] (D) to [out=-30,in=-150] (G);
\draw[->] (E) to [out=-30,in=-150] (G);
\draw[->] (F) to (G);

\node[] (Space) at (0.0,-0.5) {};
\end{tikzpicture}
\begin{tikzpicture}
\pathCompressionNodes{Result:}{PCSimplexR_Inactive}{PCSimplexR_Inactive}{PCSimplexR_Inactive}{PCSimplexR_Inactive}{PCSimplexR_Inactive}{PCSimplexR_Inactive}{highlightedSimplexVertexR}

\draw[->] (A) to [out=-25,in=-150] (G);
\draw[->] (B) to [out=-25,in=-150] (G);
\draw[->] (C) to [out=-25,in=-150] (G);
\draw[->] (D) to [out=-25,in=-150] (G);
\draw[->] (E) to [out=-25,in=-150] (G);
\draw[->] (F) to (G);
\end{tikzpicture}
\vspace*{-20pt}
    \caption{Path compression example showing vertices as circles and current vertex assignment as arrows.
    The number attached to a vertex is the order field function value, and the outer ring shows if a vertex converged (red).
    The gradient step assigns the largest neighbor to each vertex and each following Iteration sets the neighbor's neighbor for active vertices.
    Please note that the order in which every iteration is executed matters. In this example, it starts with the smallest active vertex and continues in an increasing fashion. If iteration 1 would start with the largest vertex in decreasing order the assignment would already terminate after the first iteration. \label{fig:segmentation_explenaitation}}
\end{figure}
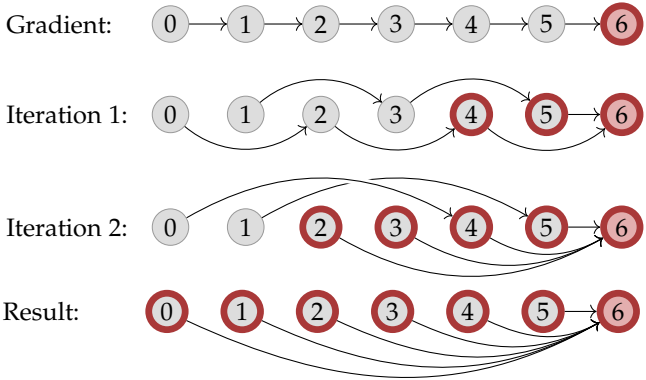

Maximum assignment for each vertex can be achieved by iteratively finding the largest neighbor's largest neighbor. As this process is lengthy, taking many steps to converge to the maximum, path compression~\cite{seidel2005top} is used to double the step size in each iteration. \autoref{fig:segmentation_explenaitation} gives an example path compression run using 7 ordered vertices.

The segmentation computation starts by assigning each vertex $v$ to its largest neighbor in the triangulation according to the order field function value $dsc(v) = argmax_{x \in N(v)} \bar{f}(x)$, allowing for a fast lookup later in the process. $N(v)$ is the set of all vertices that are connected to $v$ via an edge in the triangulation.

At the same time, maxima can be extracted by recording cases where no larger neighbor is found. For further processing, each vertex that is not a maximum is written to a list of vertices $L_0$ that did not find their maximum yet.

In the second step, the maximum for each vertex in $L_0$ is found using path compression. Here, the value of each vertex gets assigned to the largest neighbor's largest neighbor $dsc(v) = dsc(dsc(v))$. This allows doubling the step size towards the maximum in each iteration. If the corresponding maximum is not found $dsc(v) \neq dsc(dsc(v))$, the vertex did not converge to its maximum and is written to a second list $L_1$. After fully iterating over $L_0$, the process starts again using $L_1$, i.e. $L_0 = L_1$. If $L_1 = \emptyset$ after iterating over $L_0$ the maximum $dsc(v)$ is found for every vertex $v$. 

In parallel environments, each vertex can be evaluated independently with little communication in between iterations, as both steps iterate over a set of vertices. Here, the first step of finding the largest neighbor is equally distributed such that every thread executes the same amount of vertices. Still, every thread $t$ keeps a local list of active vertices, i.e. $L_{0t}, L_{1t}$, executing the following iterations independently for each thread. It is also possible to compute the ascending and descending segmentations simultaneously, further improving performance. To do this, both the largest and smallest neighbors are found at the same time in the first step, and a vertex is added to $L_1$ if it did not converge in both directions in the second step.

\subsubsection{Morse-Smale Segmentation}
The ascending and descending segmentation obtained by the algorithm above can be combined into an MS segmentation. As the descending segmentation assigns a maximum to each vertex and the ascending segmentation assigns a minimum to each vertex, vertices with the same minimum and maximum are assigned to the same MS id. Therefore, the extremum pair combination is written to each vertex as a tuple, allowing access to the involved extrema. This process is trivial to parallelize, as each thread can independently write the MS ids for its vertices.

\subsection{Multi-Label Marching Triangles/Tetrahedra}
\label{sec:marchingtet}
To visually divide MS regions from each other, region-separating geometries can be created between the regions. As the triangulation consists of triangles in the 2D case and tetrahedra in the 3D case, both cases have to be treated in slightly different ways. In 2D, region-separating geometry is created using edges that split triangles with multiple labels, whereas in 3D, triangles are utilized to separate the vertices of multi-label tetrahedra. Like marching tetrahedra~\cite{doi1991efficient}, each tetrahedron or triangle is evaluated independently, considering the labels at its vertices for generating the bisecting geometry.

\subsubsection{Triangles}
\label{sec:MarchingTriangles}
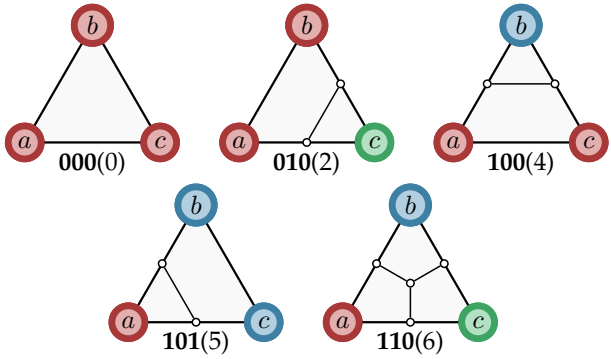
\begin{figure}[!htb]
\centering
\newcommand{\marchingTriSpace}{1em}
\newcommand{\marchingTriScaleX}{0.9}
\newcommand{\marchingTriScaleY}{0.9}
\newcommand{\marchingTriLineWidth}{0.3mm}

\newcommand{\marchingTriNodes}[6]{
    \node[#1] (A) at (0.0,0.0) {#4};
    \node[#2] (B) at (1.0,1.732) {#5};
    \node[#3] (C) at (2.0,0.0) {#6};
}

\newcommand{\marchingTriEdges}{
    \fill [simplexTriangle, draw=none, fill opacity=0.025, fill=black] (A.center) -- (B.center) -- (C.center) -- (A.center);
    \draw [line width=\marchingTriLineWidth] (A.center) -- (B.center) -- (C.center) -- (A.center) -- cycle;
}

\newcommand{\marchingTriText}[1]{
    \coordinate [label=below:{#1}] (Text) at ($(A.center)!0.5!(C.center)$);
}

\begin{tikzpicture}[xscale=\marchingTriScaleX,yscale=\marchingTriScaleY]
    \marchingTriNodes{highlightedSimplexVertexR}{highlightedSimplexVertexR}{highlightedSimplexVertexR}{$a$}{$b$}{$c$}
    
    \marchingTriText{\textbf{000}(0)}

    \marchingTriEdges

    \marchingTriNodes{highlightedSimplexVertexR}{highlightedSimplexVertexR}{highlightedSimplexVertexR}{$a$}{$b$}{$c$}
\end{tikzpicture}
\hspace{\marchingTriSpace}
\begin{tikzpicture}[xscale=\marchingTriScaleX,yscale=\marchingTriScaleY]

    \marchingTriNodes{highlightedSimplexVertexR}{highlightedSimplexVertexR}{highlightedSimplexVertexG}{$a$}{$b$}{$c$}
    
    \marchingTriText{\textbf{010}(2)}

    \marchingTriEdges
    
    \node[simplexVertex] (Q1) at (barycentric cs:A=1,C=1) {};
    \node[simplexVertex] (Q2) at (barycentric cs:B=1,C=1) {};
    \draw[thinEdge] (Q1) -- (Q2);

    \marchingTriNodes{highlightedSimplexVertexR}{highlightedSimplexVertexR}{highlightedSimplexVertexG}{$a$}{$b$}{$c$}
\end{tikzpicture}
\hspace{\marchingTriSpace}
\begin{tikzpicture}[xscale=\marchingTriScaleX,yscale=\marchingTriScaleY]

    \marchingTriNodes{highlightedSimplexVertexR}{highlightedSimplexVertexB}{highlightedSimplexVertexR}{$a$}{$b$}{$c$}
    
    \marchingTriText{\textbf{100}(4)}

    \marchingTriEdges
    
    \node[simplexVertex] (Q1) at (barycentric cs:A=1,B=1) {};
    \node[simplexVertex] (Q2) at (barycentric cs:B=1,C=1) {};
    \draw[thinEdge] (Q1) -- (Q2);

    \marchingTriNodes{highlightedSimplexVertexR}{highlightedSimplexVertexB}{highlightedSimplexVertexR}{$a$}{$b$}{$c$}
\end{tikzpicture}
\hspace{\marchingTriSpace}
\begin{tikzpicture}[xscale=\marchingTriScaleX,yscale=\marchingTriScaleY]

    \marchingTriNodes{highlightedSimplexVertexR}{highlightedSimplexVertexB}{highlightedSimplexVertexB}{$a$}{$b$}{$c$}
    
    \marchingTriText{\textbf{101}(5)}

    \marchingTriEdges
    
    \node[simplexVertex] (Q1) at (barycentric cs:A=1,B=1) {};
    \node[simplexVertex] (Q2) at (barycentric cs:A=1,C=1) {};
    \draw[thinEdge] (Q1) -- (Q2);

    \marchingTriNodes{highlightedSimplexVertexR}{highlightedSimplexVertexB}{highlightedSimplexVertexB}{$a$}{$b$}{$c$}
\end{tikzpicture}
\hspace{\marchingTriSpace}
\begin{tikzpicture}[xscale=\marchingTriScaleX,yscale=\marchingTriScaleY]

    \marchingTriNodes{highlightedSimplexVertexR}{highlightedSimplexVertexB}{highlightedSimplexVertexG}{$a$}{$b$}{$c$}
    
    \marchingTriText{\textbf{110}(6)}

    \marchingTriEdges
    
    \node[simplexVertex] (Q0) at (barycentric cs:A=1,B=1,C=1) {};
    \node[simplexVertex] (Q1) at (barycentric cs:A=1,B=1) {};
    \node[simplexVertex] (Q2) at (barycentric cs:B=1,C=1) {};
    \node[simplexVertex] (Q3) at (barycentric cs:A=1,C=1) {};
    \draw[thinEdge] (Q0) -- (Q1);
    \draw[thinEdge] (Q0) -- (Q2);
    \draw[thinEdge] (Q0) -- (Q3);

    \marchingTriNodes{highlightedSimplexVertexR}{highlightedSimplexVertexB}{highlightedSimplexVertexG}{$a$}{$b$}{$c$}
\end{tikzpicture}
\vspace*{-20pt}
\caption{All five valid cases for splitting a triangle. The label at the vertices of the triangles is drawn by color (red = $0$, blue = $1$, green = $2$), showing binary code and integer representation at the bottom, and the resulting separating edge(s) in the middle. White circles mark the points of intersection on the edges of the triangle. \label{fig:TriangleCases}}
\end{figure}

In the 2D case, a triangle can either have 1, 2, or 3 unique labels at its vertices. In the case of 1 label, no edges have to be generated as the vertices belong to the same region. When 2 different labels are present, one vertex $a$ has a different label than the other two vertices $b,c$. Here, as shown in \autoref{fig:TriangleCases}, the centers of the edges connecting $a$ to $b$ and $a$ to $c$ are used as the endpoints for the edge that splits the labels. In the case of 3 unique labels, an edge is created from the triangle center to all three of its edges.

\begin{figure}[!hb]
\centering
\begin{tikzpicture}

\node (S) at (4,1.2) {};

\node[relationship] (D1) at (2,0) {$a == c$};
\node[relationship] (D2) at (4,0) {$a == b$};
\node[relationship] (D3) at (6,0) {$a == c$};
\node[relationship] (D4) at (6,-2) {$b == c$};

\node (E0) at (0.3,0)  {\textbf{000}(0)};
\node (E1) at (2,-1.4) {\textbf{010}(2)};
\node (E2) at (7.7,0)  {\textbf{100}(4)};
\node (E3) at (7.7,-2) {\textbf{101}(5)};
\node (E4) at (4.3,-2) {\textbf{110}(6)};

\draw[->] (S) to (D2);

\draw[->] (D2) to node[above]{no} (D3);
\draw[->] (D2) to node[above]{yes} (D1);
\draw[->] (D3) to node[right]{no} (D4);

\draw[->] (D1) to node[above]{yes} (E0);
\draw[->] (D1) to node[right]{no}  (E1);
\draw[->] (D3) to node[above]{yes} (E2);
\draw[->] (D4) to node[above]{yes} (E3);
\draw[->] (D4) to node[above]{no}  (E4);

\end{tikzpicture}
\vspace*{-20pt}
\caption{Decision tree for the triangle binary code creation. Each rhombus represents a decision, each node shows the resulting code with the bit representation in bold and integer representation in brackets.\label{fig:TriangleDescisionTree}}
\end{figure}
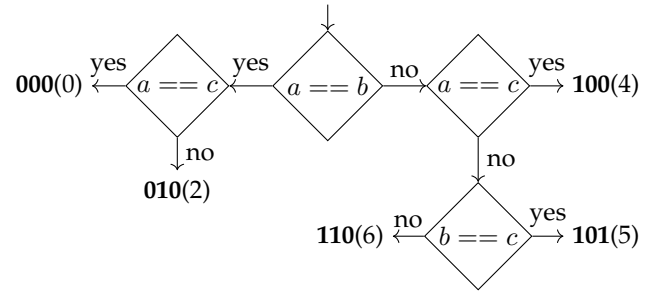

Computationally, this is achieved using a lookup table that describes every possible configuration in a triangle. Therefore, a 3-bit binary code with values in the range of $\{0, .., 6\}$ is created to describe the current triangle configuration, utilizing the labels at the three vertices $a, b, c$ of the considered triangle, converted to a dense local representation such that $a = 0, b \in \{0, 1\}, c \in \{0, 1, 2\}$. The label $a$ is always considered to be $0$, $b$ can either be $0$ or $1$ depending on the equality to label $a$, and $c$ can either be $0$, $1$, or $2$ depending on the equality to labels $a$ and $b$. Therefore, $b \in \{0,1\}$ determines the left bit and $c \in \{0,1,2\}$ determines the last two bits. \autoref{fig:TriangleDescisionTree} provides the decision tree. It should be noted that some binary codes cannot appear, as $c = 1$ can only be the case if $b = 1$. Also, codes like $011(3)$ are not possible in general, as there is no label $3$.

Each of the five valid binary codes corresponds to a triangle configuration, as shown in \autoref{fig:TriangleCases}. This allows retrieving the triangle edges that need to be connected. The whole procedure is well-scaling as it is executed per triangle.

\subsubsection{Tetrahedra}
In 3D, the domain is subdivided into tetrahedra. Therefore, up to 4 unique labels $a, b, c, d$ can be present at the vertices of a tetrahedron. Similar to the triangle case, a 5-bit binary code is created and translated into a tetrahedron configuration. This configuration is used to create a consistent triangulation that separates unique labels from each other.

The binary code for tetrahedra is created by an extended logic. $a$ is considered to be 0 again, $b \in \{0,1\}$ determines the left bit, $c \in \{0,1,2\}$ determines the next 2 bits and $d \in \{0,1,2,3\}$ determines the last two bits. If the label of a vertex with lower index matches, its label is used as the resulting label, otherwise the index of the own vertex is used. All valid configurations are provided in \autoref{tab:TetBinaryCode}. Some binary codes are invalid, as some labels might not exist and can not be assigned to a vertex of a higher index. E.g. $00010(2)$ is not possible as $d$ would have label $2$, but $c$ has label $0$, making label $2$ non-existent in this configuration.

\begin{table}[!htb]
    \caption{All valid binary codes, the codes converted to an integer, the number of unique labels of the tetrahedron, and the value of each label. \label{tab:TetBinaryCode}}
    \centering
    \begin{tabular}{|c|c|c|c|c|c|c|}
    \hline
        Binary code  & Case & \#Labels & $a$ & $b$ & $c$ & $d$ \\
        \hline
        00000 & 0  & 1 & 0 & 0 & 0 & 0 \\
        00011 & 3  & 2 & 0 & 0 & 0 & 3 \\
        01000 & 8  & 2 & 0 & 0 & 2 & 0 \\
        01010 & 10 & 2 & 0 & 0 & 2 & 2 \\
        01011 & 11 & 3 & 0 & 0 & 2 & 3 \\
        10000 & 16 & 2 & 0 & 1 & 0 & 0 \\
        10001 & 17 & 2 & 0 & 1 & 0 & 1 \\
        10011 & 19 & 3 & 0 & 1 & 0 & 3 \\
        10100 & 20 & 2 & 0 & 1 & 1 & 0 \\
        10101 & 21 & 2 & 0 & 1 & 1 & 1 \\
        10111 & 23 & 3 & 0 & 1 & 1 & 3 \\
        11000 & 24 & 3 & 0 & 1 & 2 & 0 \\
        11001 & 25 & 3 & 0 & 1 & 2 & 1 \\
        11010 & 26 & 3 & 0 & 1 & 2 & 2 \\
        11011 & 27 & 4 & 0 & 1 & 2 & 3 \\
        \hline
    \end{tabular}
\end{table}

\begin{figure}[!htb]
    \centering
    \newcommand{\marchingTetrahedraSpace}{2em}
\newcommand{\marchingTetrahedraScaleX}{3}
\newcommand{\marchingTetrahedraScaleY}{2.75}
\newcommand{\marchingTetrahedraLineWidth}{0.3mm}
\pgfmathsetmacro{\factor}{1/sqrt(2)}

\newcommand{\marchingTetrahedraNodes}[8]{
    \node[#1] (A) at (0.0,0.1) {#5};
    \node[#2] (B) at (0.9,0.0) {#6};
    \node[#3] (C) at (0.2,-0.5) {#7};
    \node[#4] (D) at (0.3,0.4) {#8};
}

\newcommand{\marchingTetrahedraEdges}{
    \fill [simplexTriangle, draw=none, fill opacity=0.025, fill=black] (A.center) -- (C.center) -- (D.center) -- (A.center);
    \fill [simplexTriangle, draw=none, fill opacity=0.1, fill=black] (B.center) -- (C.center) -- (D.center) -- (B.center);
    \draw [line width=\marchingTetrahedraLineWidth, dashed, opacity=0.5] (A.center) -- (B.center);
    \draw [line width=\marchingTetrahedraLineWidth] (A.center) -- (D.center) -- (B.center) -- (C.center) -- cycle;
    \draw [line width=\marchingTetrahedraLineWidth] (C.center) -- (D.center);
}

\newcommand{\marchingTetrahedraText}[1]{
    \coordinate [label=right:{Cases: #1}] (Text) at (0.0,-0.7);
}

\begin{tikzpicture}[xscale=\marchingTetrahedraScaleX,yscale=\marchingTetrahedraScaleY]

    \marchingTetrahedraNodes{highlightedSimplexVertexR}{highlightedSimplexVertexB}{highlightedSimplexVertexB}{highlightedSimplexVertexB}{$a$}{$b$}{$c$}{$d$}
    
    \marchingTetrahedraText{3, 8, 16, 21}

    \marchingTetrahedraEdges
    
    \node[simplexVertex] (Q1) at (barycentric cs:A=1,B=1) {};
    \node[simplexVertex] (Q2) at (barycentric cs:A=1,C=1) {};
    \node[simplexVertex] (Q3) at (barycentric cs:A=1,D=1) {};
    
    \fill [highlightedSimplexTriangleB, fill opacity=0.45, thinEdge] (Q1.center) -- (Q2.center) -- (Q3.center) -- cycle;
    
    \node[simplexVertex] (Q1) at (barycentric cs:A=1,B=1) {};
    \node[simplexVertex] (Q2) at (barycentric cs:A=1,C=1) {};
    \node[simplexVertex] (Q3) at (barycentric cs:A=1,D=1) {};

    \marchingTetrahedraNodes{highlightedSimplexVertexR}{highlightedSimplexVertexB}{highlightedSimplexVertexB}{highlightedSimplexVertexB}{$a$}{$b$}{$c$}{$d$}
\end{tikzpicture}
\hspace{\marchingTetrahedraSpace}
\begin{tikzpicture}[xscale=\marchingTetrahedraScaleX,yscale=\marchingTetrahedraScaleY]

    \marchingTetrahedraNodes{highlightedSimplexVertexR}{highlightedSimplexVertexR}{highlightedSimplexVertexB}{highlightedSimplexVertexB}{$a$}{$b$}{$c$}{$d$}
    
    \marchingTetrahedraText{10, 17, 20}

    \marchingTetrahedraEdges
    
    \node[simplexVertex] (Q1) at (barycentric cs:A=1,C=1) {};
    \node[simplexVertex] (Q2) at (barycentric cs:A=1,D=1) {};
    \node[simplexVertex] (Q3) at (barycentric cs:B=1,C=1) {};
    \node[simplexVertex] (Q4) at (barycentric cs:B=1,D=1) {};
    
    \fill [highlightedSimplexTriangleB, fill opacity=0.45, thinEdge] (Q1.center) -- (Q2.center) -- (Q3.center) -- cycle;
    \fill [highlightedSimplexTriangleB, fill opacity=0.45, thinEdge] (Q2.center) -- (Q3.center) -- (Q4.center) -- cycle;
    
    \node[simplexVertex] (Q1) at (barycentric cs:A=1,C=1) {};
    \node[simplexVertex] (Q2) at (barycentric cs:A=1,D=1) {};
    \node[simplexVertex] (Q3) at (barycentric cs:B=1,C=1) {};
    \node[simplexVertex] (Q4) at (barycentric cs:B=1,D=1) {};

    \marchingTetrahedraNodes{highlightedSimplexVertexR}{highlightedSimplexVertexR}{highlightedSimplexVertexB}{highlightedSimplexVertexB}{$a$}{$b$}{$c$}{$d$}
\end{tikzpicture}
\hspace{\marchingTetrahedraSpace}
\begin{tikzpicture}[xscale=\marchingTetrahedraScaleX,yscale=\marchingTetrahedraScaleY]

    \marchingTetrahedraNodes{highlightedSimplexVertexR}{highlightedSimplexVertexR}{highlightedSimplexVertexB}{highlightedSimplexVertexG}{$a$}{$b$}{$c$}{$d$}
    
    \marchingTetrahedraText{11, 19, 23-26}

    \marchingTetrahedraEdges
    
    \node[simplexVertex] (T1) at (barycentric cs:A=1,C=1,D=1) {};
    \node[simplexVertex] (T2) at (barycentric cs:B=1,C=1,D=1) {};
    \node[simplexVertex] (E1) at (barycentric cs:A=1,C=1) {};
    \node[simplexVertex] (E2) at (barycentric cs:B=1,C=1) {};
    \node[simplexVertex] (E3) at (barycentric cs:A=1,D=1) {};
    \node[simplexVertex] (E4) at (barycentric cs:B=1,D=1) {};
    \node[simplexVertex] (E5) at (barycentric cs:C=1,D=1) {};
    
    \fill [highlightedSimplexTriangleB, fill opacity=0.45, thinEdge] (T1.center) -- (T2.center) -- (E1.center) -- cycle;
    \fill [highlightedSimplexTriangleB, fill opacity=0.45, thinEdge] (T2.center) -- (E1.center) -- (E2.center) -- cycle;
    \fill [highlightedSimplexTriangleB, fill opacity=0.45, thinEdge] (T1.center) -- (T2.center) -- (E5.center) -- cycle;
    \fill [highlightedSimplexTriangleB, fill opacity=0.45, thinEdge] (T1.center) -- (T2.center) -- (E3.center) -- cycle;
    \fill [highlightedSimplexTriangleB, fill opacity=0.45, thinEdge] (T2.center) -- (E3.center) -- (E4.center) -- cycle;
    
    \node[simplexVertex] (T1) at (barycentric cs:A=1,C=1,D=1) {};
    \node[simplexVertex] (T2) at (barycentric cs:B=1,C=1,D=1) {};
    \node[simplexVertex] (E1) at (barycentric cs:A=1,C=1) {};
    \node[simplexVertex] (E2) at (barycentric cs:B=1,C=1) {};
    \node[simplexVertex] (E3) at (barycentric cs:A=1,D=1) {};
    \node[simplexVertex] (E4) at (barycentric cs:B=1,D=1) {};
    \node[simplexVertex] (E5) at (barycentric cs:C=1,D=1) {};

    \marchingTetrahedraNodes{highlightedSimplexVertexR}{highlightedSimplexVertexR}{highlightedSimplexVertexB}{highlightedSimplexVertexG}{$a$}{$b$}{$c$}{$d$}
\end{tikzpicture}
\hspace{\marchingTetrahedraSpace}
\begin{tikzpicture}[xscale=\marchingTetrahedraScaleX,yscale=\marchingTetrahedraScaleY]

    \marchingTetrahedraNodes{highlightedSimplexVertexR}{highlightedSimplexVertexY}{highlightedSimplexVertexB}{highlightedSimplexVertexG}{$a$}{$b$}{$c$}{$d$}
    
    \marchingTetrahedraText{27}

    \marchingTetrahedraEdges
    
    \node[simplexVertex] (E0) at (barycentric cs:A=1,B=1) {};
    \node[simplexVertex] (E1) at (barycentric cs:A=1,C=1) {};
    \node[simplexVertex] (E2) at (barycentric cs:A=1,D=1) {};
    \node[simplexVertex] (E3) at (barycentric cs:B=1,C=1) {};
    \node[simplexVertex] (E4) at (barycentric cs:B=1,D=1) {};
    \node[simplexVertex] (E5) at (barycentric cs:C=1,D=1) {};
    \node[simplexVertex] (T0) at (barycentric cs:A=1,B=1,C=1) {};
    \node[simplexVertex] (T1) at (barycentric cs:A=1,B=1,D=1) {};
    \node[simplexVertex] (T2) at (barycentric cs:A=1,C=1,D=1) {};
    \node[simplexVertex] (T3) at (barycentric cs:B=1,C=1,D=1) {};
    \node[simplexVertex] (C0) at (barycentric cs:A=1,B=1,C=1,D=1) {};
    
    \fill [highlightedSimplexTriangleB, fill opacity=0.45, thinEdge] (C0.center) -- (T0.center) -- (E0.center) -- cycle;
    \fill [highlightedSimplexTriangleB, fill opacity=0.45, thinEdge] (C0.center) -- (T0.center) -- (E1.center) -- cycle;
    \fill [highlightedSimplexTriangleB, fill opacity=0.45, thinEdge] (C0.center) -- (T0.center) -- (E3.center) -- cycle;
    
    \fill [highlightedSimplexTriangleB, fill opacity=0.45, thinEdge] (C0.center) -- (T1.center) -- (E0.center) -- cycle;
    \fill [highlightedSimplexTriangleB, fill opacity=0.45, thinEdge] (C0.center) -- (T1.center) -- (E2.center) -- cycle;
    \fill [highlightedSimplexTriangleB, fill opacity=0.45, thinEdge] (C0.center) -- (T1.center) -- (E4.center) -- cycle;
    
    \fill [highlightedSimplexTriangleB, fill opacity=0.45, thinEdge] (C0.center) -- (T2.center) -- (E1.center) -- cycle;
    \fill [highlightedSimplexTriangleB, fill opacity=0.45, thinEdge] (C0.center) -- (T2.center) -- (E2.center) -- cycle;
    \fill [highlightedSimplexTriangleB, fill opacity=0.45, thinEdge] (C0.center) -- (T2.center) -- (E5.center) -- cycle;

    \fill [highlightedSimplexTriangleB, fill opacity=0.45, thinEdge] (C0.center) -- (T3.center) -- (E3.center) -- cycle;
    \fill [highlightedSimplexTriangleB, fill opacity=0.45, thinEdge] (C0.center) -- (T3.center) -- (E4.center) -- cycle;
    \fill [highlightedSimplexTriangleB, fill opacity=0.45, thinEdge] (C0.center) -- (T3.center) -- (E5.center) -- cycle;
    
    \node[simplexVertex] (E0) at (barycentric cs:A=1,B=1) {};
    \node[simplexVertex] (E1) at (barycentric cs:A=1,C=1) {};
    \node[simplexVertex] (E2) at (barycentric cs:A=1,D=1) {};
    \node[simplexVertex] (E3) at (barycentric cs:B=1,C=1) {};
    \node[simplexVertex] (E4) at (barycentric cs:B=1,D=1) {};
    \node[simplexVertex] (E5) at (barycentric cs:C=1,D=1) {};
    \node[simplexVertex] (T0) at (barycentric cs:A=1,B=1,C=1) {};
    \node[simplexVertex] (T1) at (barycentric cs:A=1,B=1,D=1) {};
    \node[simplexVertex] (T2) at (barycentric cs:A=1,C=1,D=1) {};
    \node[simplexVertex] (T3) at (barycentric cs:B=1,C=1,D=1) {};
    \node[simplexVertex] (C0) at (barycentric cs:A=1,B=1,C=1,D=1) {};

    \marchingTetrahedraNodes{highlightedSimplexVertexR}{highlightedSimplexVertexY}{highlightedSimplexVertexB}{highlightedSimplexVertexG}{$a$}{$b$}{$c$}{$d$}
\end{tikzpicture}
\vspace*{-20pt}
    \caption{Triangulation for tetrahedra with more than 1 unique label ignoring permutations and rotations. The case number below the tetrahedra refers to \autoref{tab:TetBinaryCode}. \label{fig:tetrahedraTriangulation}}
\end{figure}
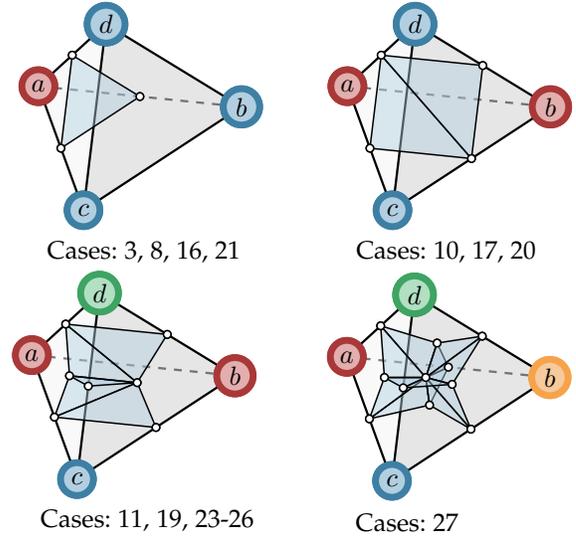

After the binary code is determined, a lookup table is used to retrieve the resulting separating triangles utilizing the edge, triangle, and tetrahedra centers to be connected. This allows for a fast triangulation of the tetrahedra, as the resulting triangle vertices are directly retrieved. \autoref{fig:tetrahedraTriangulation} shows the resulting triangulation for all cases ignoring permutations and rotations. The triangulation across tetrahedra is always consistent, as the triangle labels of the tetrahedron mimic the 2D case, i.e. triangles are either split by connecting their edge centers to their triangle center, or two triangle edges are connected. Therefore, the triangle connecting two incident tetrahedra is always split in the same way and the resulting triangulation separates labels from each other without any holes in the geometry.

\subsubsection{Triangulation Options}
As the information requested slightly varies from user to user, two region-separating geometries are available. The user can utilize any segmentation (ascending, descending, and Morse-Smale) together with either region boundaries or region separators. \autoref{fig:wallmodes} showcases a closeup view of the segmentations (b-e), where (b) and (d) show region separators, and (c) and (e) show region boundaries.

\noindent\textbf{Segmentation Selection:}
For some applications, a certain segmentation will be of great interest. Here, either the MS, ascending, or descending segmentation can be chosen to deliver the labels for the marching tetrahedra algorithm.  Instead of using the MS segmentation, it is also possible to show the union of the ascending and descending segmentation, producing intersecting geometry at the meeting points of both region-separating geometries.

\noindent\textbf{Region Separating Geometries:}
Both the region separators and the region boundaries have slightly different use cases and allow to map different information to its geometry. The region separators divide all regions, or areas of influence of minima-maxima pairs, from each other by building geometry between them. Therefore, information about the minima and maxima involved in each separating triangle can be displayed to the user. Each subset of the surface that separates the same two regions can be extracted here. Still, some overhead is involved in computing the region separators, as many triangles have to be used to separate the regions from each other, depicted in \autoref{fig:tetrahedraTriangulation}.

Region boundaries allow extracting the hull of regions or areas of influence of minima-maxima pairs. They use all tetrahedra with 3 vertices of the same label. Those 3 vertices form a triangle in the input simplicial complex that can be directly used as a separating geometry. This option will result in faster computation times and allows the extraction of geometry for each region.

\section{Results}
\label{sec:Results}
In this section, three example datasets and their region-separating geometry outputs for both algorithms are provided. The Noisy Terrain dataset in \autoref{sec:noisy_terrain} is used to highlight differences between both algorithms in the 2D case. \autoref{sec:at_dataset} shows that both algorithms produce a very similar output when no saddle-saddle 2-cells are present. The last two datasets indicate that the MS complex is providing more geometry than necessary to effectively extract useful information in many cases, while the PLMSS can extract the areas of influence without additional preprocessing. 


\subsection{Noisy Terrain}
\label{sec:noisy_terrain}
\begin{figure}[!hbt]
\centering
    \subfloat[PLMSS]{\label{fig:noisymss}\includegraphics[width=0.45\linewidth]{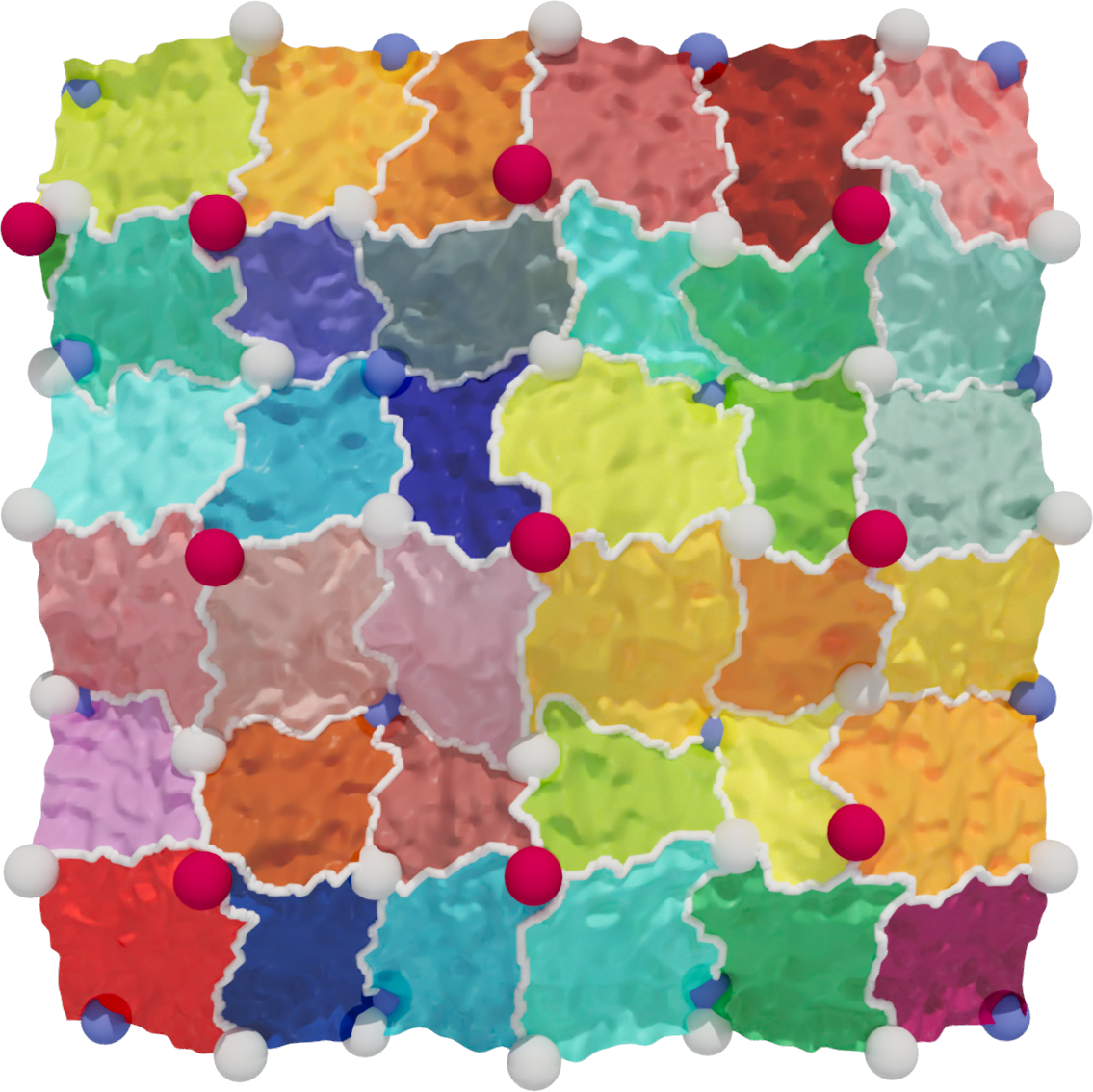}}
    \hfil
    \subfloat[MS complex]{\label{fig:noisymsc}\includegraphics[width=0.45\linewidth]{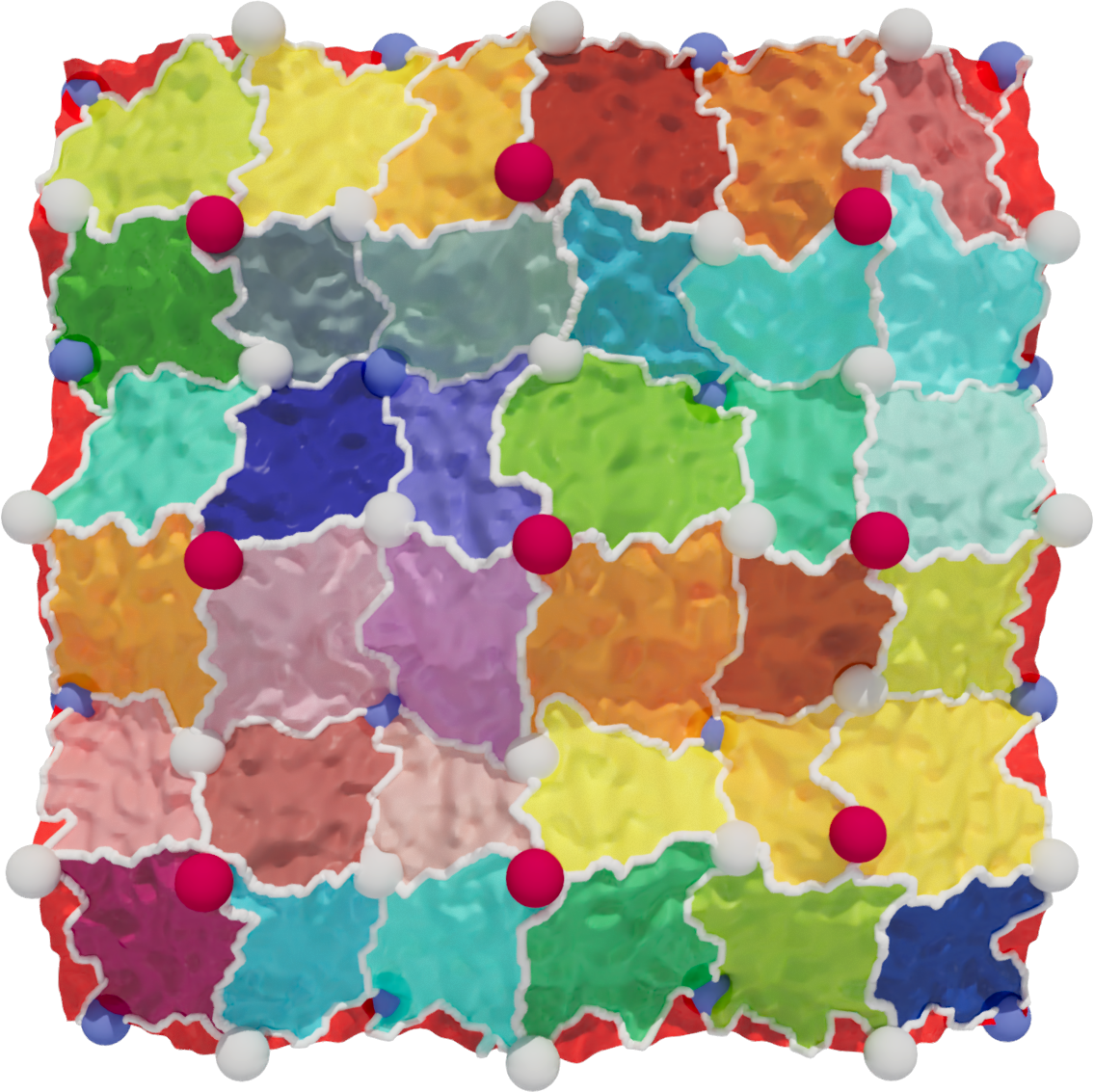}}
    \hfil
    \subfloat[Region separators]{\label{fig:noisymssmsc}\includegraphics[width=0.80\linewidth]{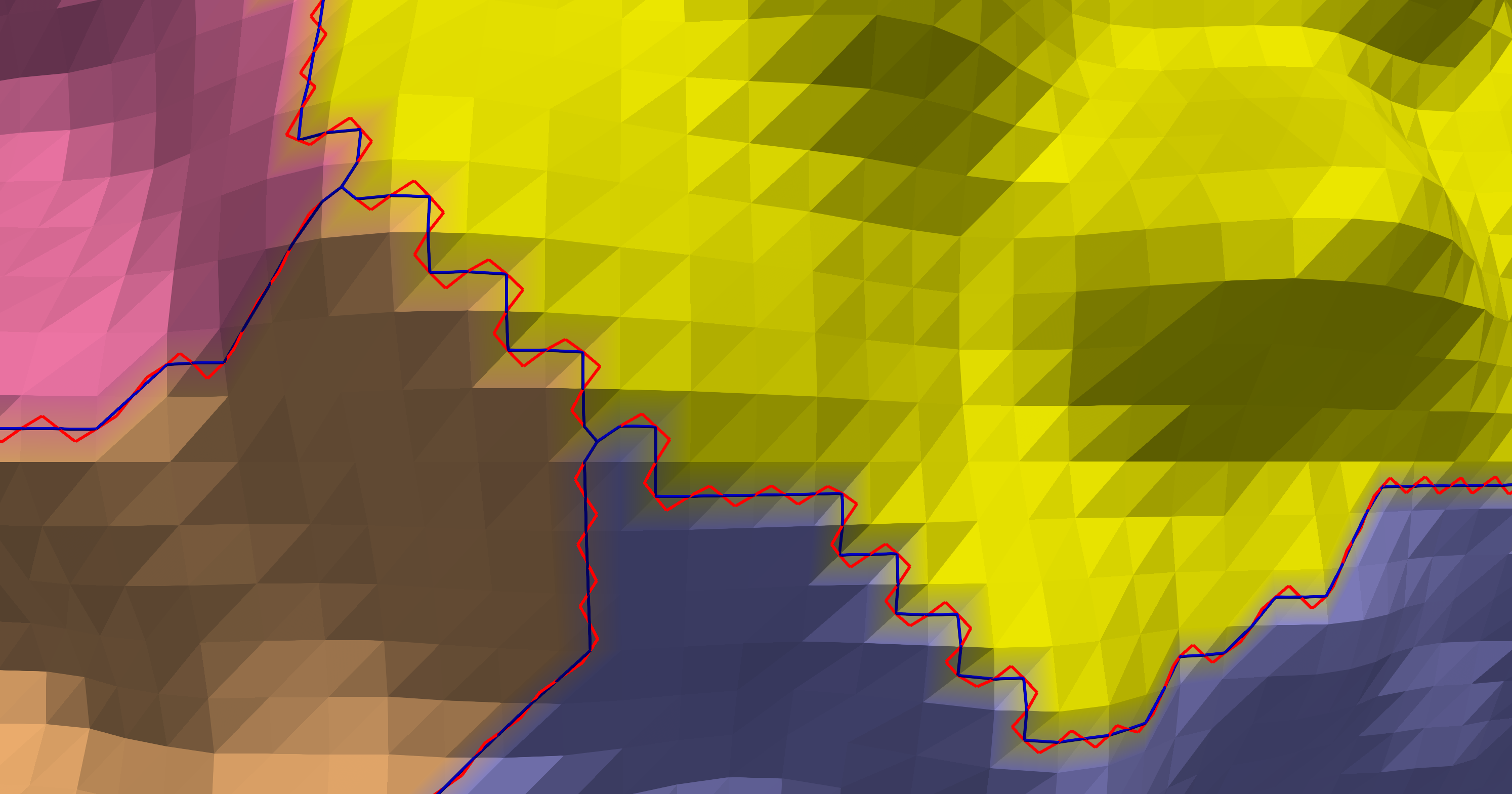}}
\caption{Noisy Terrain dataset~\cite{ttkData} showing the PLMSS and MS complex next to each other. Both show critical points (blue = Minima, white = Saddles, red = Maxima), separating geometries in white, and the surface colored by MS complex ids. The border region of the MS complex is not fully segmented 
as a vanilla implementation of the expansion-based discrete gradient computation algorithm \cite{robins2011theory} may miss maxima on the boundary,
whereas the boundary of the PLMSS is fully segmented. (c) visually compares the region-separating geometry of the PLMSS (blue) and the MS complex (red). The MS complex produces geometries with a characteristic step-function shape (red), whereas the PLMSS produces separating geometries with a linear-slope shape (blue).
\label{fig:noisyTerrain}}
\end{figure}

The Noisy Terrain dataset is a triangulated surface with elevation scalars attached, chosen as an illustrative example. It has a 300x300 resolution and can be found in the TTK data repository~\cite{ttkData}. Generally, the dataset consists of hills and valleys on a regular grid, where the hills are getting smaller the closer they are to the border. Additionally, noise was added to the terrain to showcase topological simplification.

\autoref{fig:noisyTerrain} compares the PLMSS with the MS complex, using the same color coding for critical points, separating geometries, and segmentation in both versions. Slight differences can be detected regarding the separating geometries, where the MS complex separating geometries are defined on the dual graph and are connecting triangle centers in the 2D case. Concerning the PLMSS, triangles are split according to the labels at the triangle vertices. Another difference can be spotted at the borders of the MS complex, where great sections of the border are not labeled, as a vanilla implementation of the expansion-based discrete gradient computation algorithm of Robins et al.~\cite{robins2011theory} (implemented in TTK) may miss PL maxima on the domain boundary (local post-processing of the discrete gradient would be required to enforce the detection of discrete maxima 
in the star of boundary PL maxima).
As the PLMSS assigns a maximum-minimum pair to each vertex, every vertex is properly labeled without skipping the boundary region.

\subsection{AT Molecule}
\label{sec:at_dataset}

\begin{figure}[!htb]
\centering
    \subfloat[PLMSS]{\label{fig:at_mss}\includegraphics[width=0.95\linewidth, angle=90]{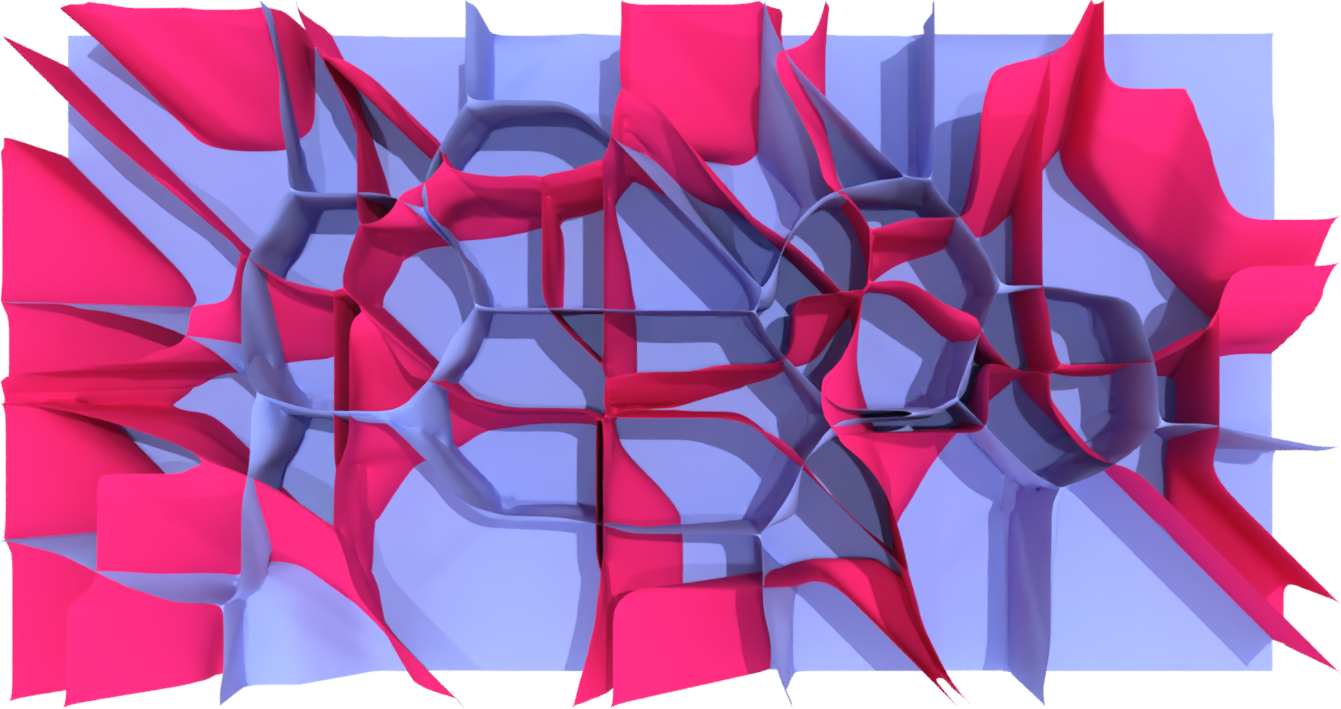}}
    \subfloat[MS complex]{\label{fig:at_MS complex}\includegraphics[width=0.95\linewidth, angle=90]{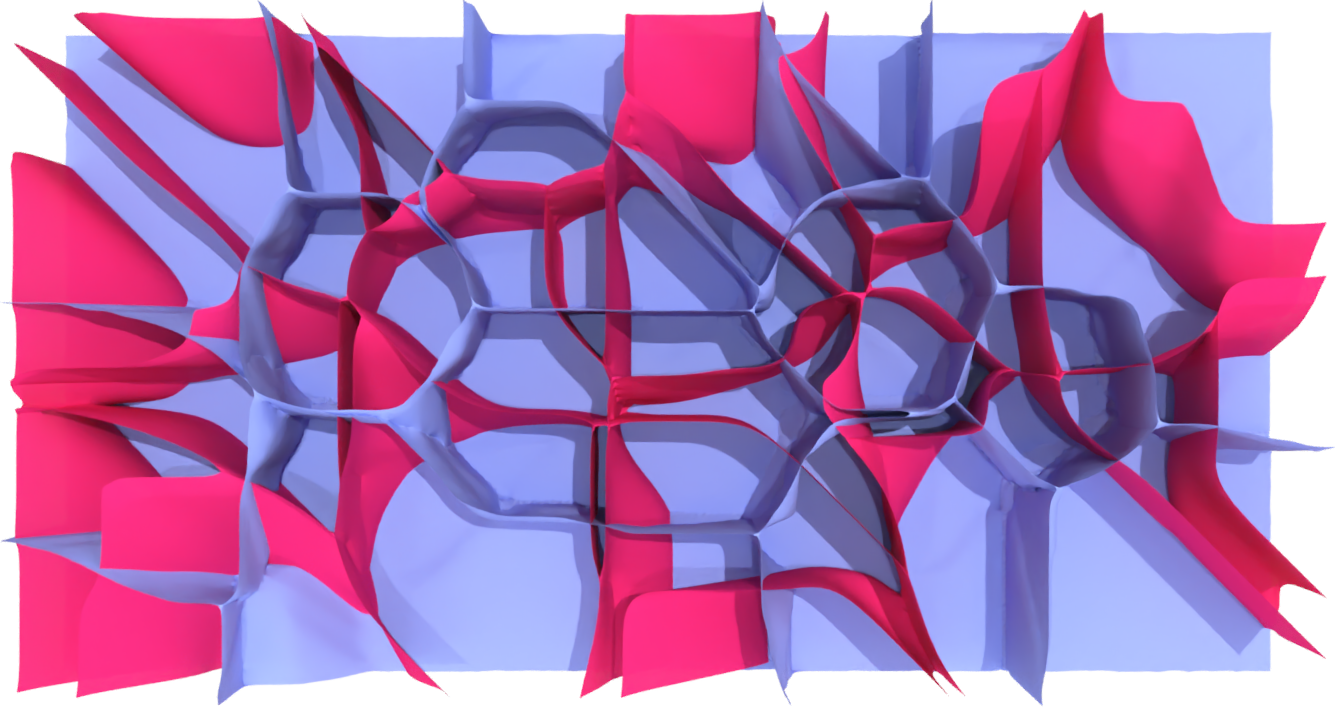}} 
\caption{Comparison between the region separators computed by PLMSS~(a), and the MS complex 2-cells computed by TTK~(b) using the AT dataset.
The resulting surfaces are colored by the segmentation type, i.e., red for the descending and blue for the ascending segmentation.
To ease visual comparison, both surfaces were smoothed 20 times using TTK's geometry smoother.
Both surfaces are almost visually identical.
}
\label{fig:at_vti}
\end{figure}

The AT dataset from the TTK Tutorial Data~\cite{ttkTutorialData} shows the simulation of the electron density of a molecule restricted to a plane but embedded in 3D space. This example, provided in \autoref{fig:at_vti}, shows that the PLMSS and MS complex extract the same underlying geometry at heart if no saddle-saddle 2-cells are present in the dataset. To allow an easier comparison of the separating geometries, all geometries were colored by the underlying segmentation (ascending in red and descending in blue) and smoothed. Most of the separating geometry coincides with at most 1 tetrahedron space in between both representations. The largest difference can be seen in the two narrow red geometries on the middle right of the images, as their distance is smaller when using the PLMSS. The difference between both representations themselves is a result of the dual graph definition of MS complex, compared to the per-vertex definition of the PLMSS.

\subsection{Viscous Fingering}
\label{sec:viscous_fingering}

\begin{figure}[tb]
\centering
    \subfloat[PLMSS]{\label{fig:viscous_mss}\includegraphics[width=0.5\linewidth]{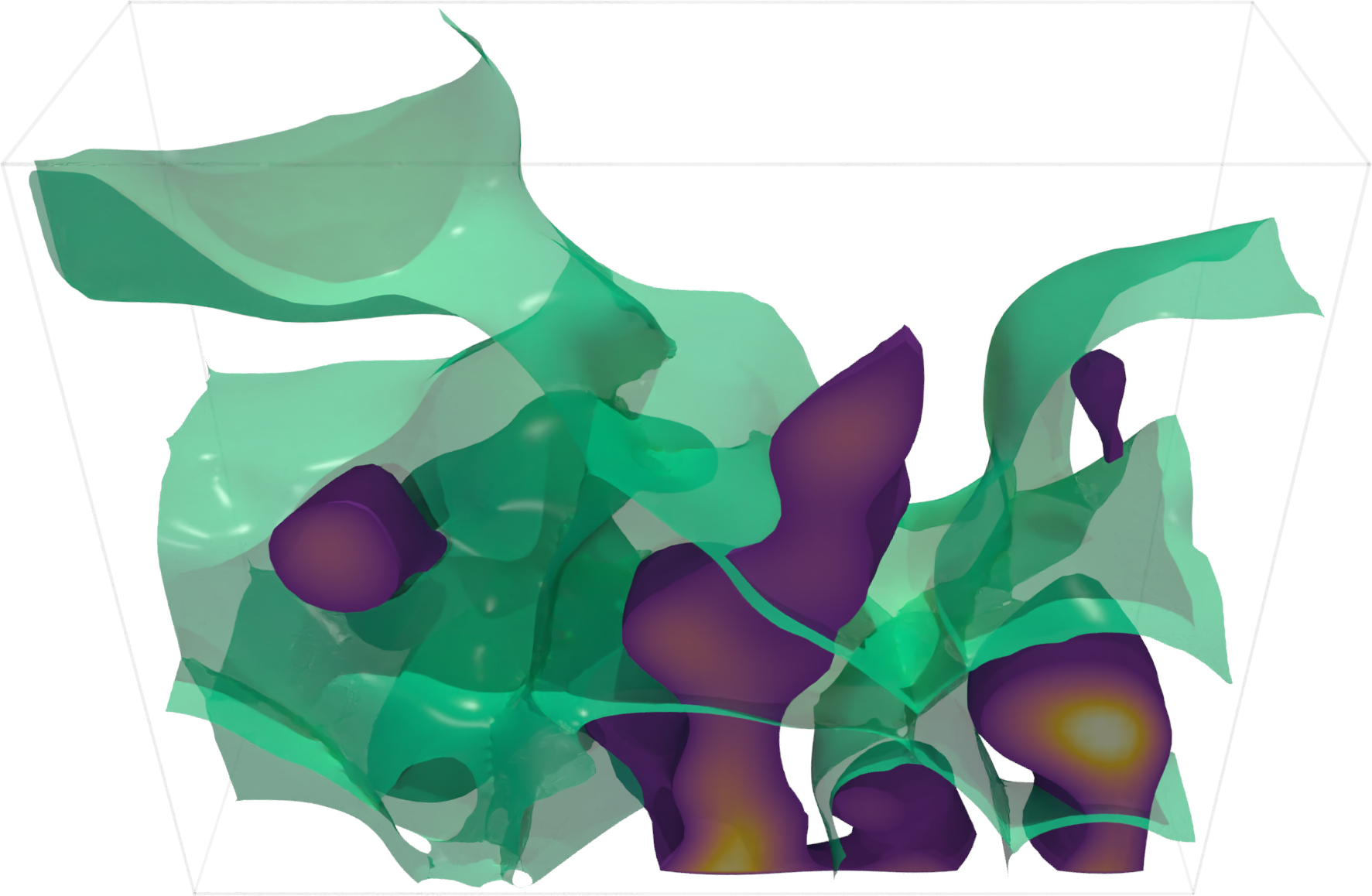}}
    \hfill
    \subfloat[MS complex]{\label{fig:viscous_msc_post}\includegraphics[width=0.5\linewidth]{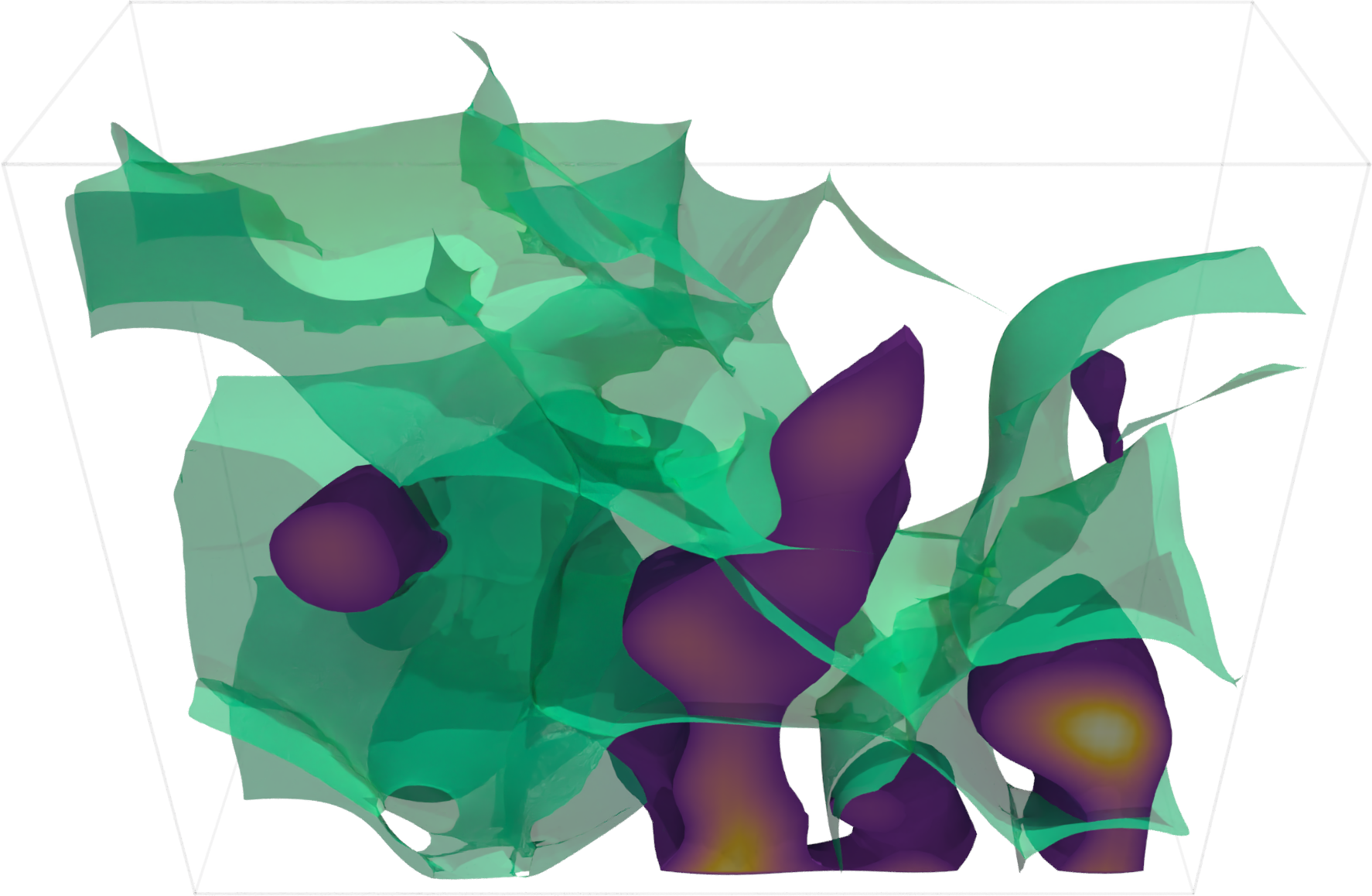}}
 \caption{Comparison of the PLMSS region boundaries and the MS complex using the Viscous Fingering dataset~\cite{ttkData} simplified with an absolute persistence threshold of $0.1$. Both images show the boundary interface of viscous fingers as contours of the salt concentration density scalar field, colored by the density from yellow (high concentration) to purple (low concentration). (a) shows that the PLMSS region boundaries can extract the region-separating geometries that separate single viscous fingers effectively without cluttering the visualization. (b) provides the original MS complex, where more geometry is extracted due to the saddle-saddle 2-cells, cluttering the image. Those saddle-saddle walls would have to be removed by additional postprocessing.\label{fig:viscous_dataset}}
\end{figure}

The Viscous Fingering dataset~\cite{ttkData} represents the result of finite pointset method simulations that simulate the mixing of salt solutions inside water. Regions of high salt concentration form structures called viscous fingers. The analysis of such structures usually involves Reeb graphs or iso-surfaces to identify single fingers in the dataset~\cite{lukasczyk2017viscous}. The MS complex suffers from additional saddle-saddle 2-cells in such scenarios that stem from discrete Morse theory, complicating effective analysis. 
Filtering these saddle-saddle 2-cells out of the set of 2-cells is usually expensive as further post-processing is required to identify and simplify the saddles responsible for such 2-cells. 
The PLMSS, on the other hand, segments the data into regions of influence of maxima-minima pairs, providing a region for each partial viscous finger. The granularity of these separating geometries can be controlled by topological simplification, allowing users to achieve the desired level of detail of the segmentation.

\autoref{fig:viscous_dataset} shows the Viscous Fingering dataset from the TTK data repository~\cite{ttkData} with an applied persistence threshold of 0.1. The colored isosurfaces provide the finger surfaces that correspond to regions of high salt concentration. To allow a deeper look into the dataset, it was clipped in the middle after all geometries were created. Figure~\ref{fig:viscous_dataset}a shows that the PLMSS effectively separates fingers from each other, providing the area of influence of the maxima and hence, for the viscous fingers. The region separators are used to extract the area of influence of single fingers to analyze the area they are growing into with increasing salt concentration. \autoref{fig:viscous_dataset}b provides the MS complex. As for additional saddle-saddle 2-cells, the image is less clear and more cluttered, which can be problematic with noisy or large datasets. Those saddle-saddle walls would have to be removed by additional postprocessing.

\subsection{Rayleigh-Taylor instability (Miranda)}
\label{sec:rayleigh-taylor}

\begin{figure*}[!bth]
  \centering
  \includegraphics[width=.95\linewidth]{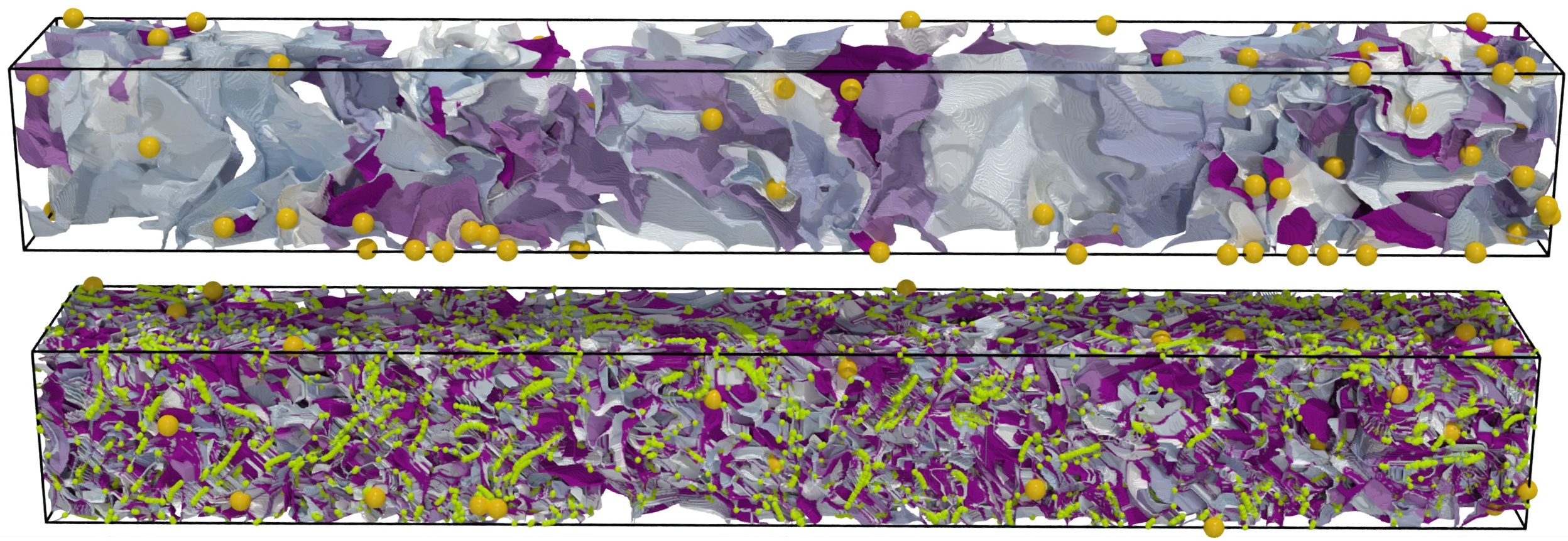}
      \caption{Illustration showing the results of the piecewise linear Morse-Smale segmentation (PLMSS) ~(top) and the Morse-Smale (MS) complex (bottom) utilizing a Rayleigh-Taylor instability simulation~\cite{miranda} dataset.  The image shows extrema as large orange spheres, additionally providing saddles as green small spheres in the MS complex, as saddle-saddle separatrices are the actual reason for the cluttered MS complex visualization. Even though filtering out saddle-saddle separatrices is possible, the computational overhead will favor the PLMSS in situations where saddle-saddle separatrices hide important features. 
      \label{fig:teaser}}
\end{figure*}

A concept of controlled fusion using hydrogen isotopes in a laser-lighted fuel capsule lead to the discovery of Rayleigh-Taylor instability\cite{miranda} at the boundary of the capsule. The simulation models the heating process of two hydrogen isotopes for fusion burn. The energy from the laser is non-uniformly distributed and causes small perturbations that quickly grow. One time step of a simulation is analyzed using the MS complex of TTK and the PLMSS.

To filter noise from the data, an absolute persistence threshold of $0.1$ was applied by utilizing localized topological simplification~\cite{lukasczyk2020localized}. By solely extracting the border surfaces of the area of influence created by maximum-minimum pairs, the PLMSS removes clutter from the separating geometries of the MS complex due to the missing saddle-saddle 2-cells. 

\autoref{fig:teaser} compares the visual results of the PLMSS with the MS complex, using one timestep of a Rayleigh-Taylor instability simulation. Here, the PLMSS manages to extract the area of influence of minima and maxima in the dataset. The region-separating geometry is very structured and the boundary between regions of interest can be identified. In contrast, the MS complex introduces a lot of noise due to the remaining saddle-saddle 2-cells that clutter the resulting image. Additionally, the maxima on the boundary are missing, 
as a vanilla implementation of the expansion-based discrete gradient computation algorithm of Robins et al.~\cite{robins2011theory} (implemented in TTK) may miss PL maxima on the domain boundary.


\section{Performance}
\label{sec:performance}

\begin{table*}[th]
    \centering
    \caption{Raw timing data of the MSCEER, TTK, and PLMSS algorithms in seconds. For each timing, the test was executed 10 times, removing the best and worst time regarding the time listed in the last row, and averaging the remaining runs. It should be noted that the top-level cell count is 6 times higher with TTK and PLMSS than with MSCEER. Similarly, the total number of cells in the input simplicial complex are differing by a factor of roughly 3.24 (Miranda MSCEER: 1,070,599,167 / TTK, PLMSS: 3,473,956,851) (Foot MSCEER: 133,432,831 / TTK, PLMSS: 432,287,731).}
    \begin{tttabular}{llrrrrrrrrrrrr}
&  & \multicolumn{6}{c}{Miranda $512^3$} & \multicolumn{6}{c}{Foot $256^3$} \\
Task & \multicolumn{1}{|c}{Algo.} & \multicolumn{1}{|r}{1T} & 2T & 4T & 8T & 16T & 24T & \multicolumn{1}{|r}{1T} & 2T & 4T & 8T & 16T & 24T \\ \hline \hline
\multirow{3}{*}{DGF} & \multicolumn{1}{|r}{{\cellcolor{mygray2}{TTK}}} & \multicolumn{1}{|r}{ {\cellcolor{mygray2}{1,326.76}}} & {\cellcolor{mygray2}{685.52}} & {\cellcolor{mygray2}{365.00}} & {\cellcolor{mygray2}{200.15}} & {\cellcolor{mygray2}{112.10}} & {\cellcolor{mygray2}{84.55}} & \multicolumn{1}{|r}{{\cellcolor{mygray2}{152.65}}} & {\cellcolor{mygray2}{78.00}} & {\cellcolor{mygray2}{42.45}} & {\cellcolor{mygray2}{23.32}} & {\cellcolor{mygray2}{13.32}} & {\cellcolor{mygray2}{10.08}}\\
& \multicolumn{1}{|r}{{\cellcolor{mygray1}{MSCEER}}} & \multicolumn{1}{|r}{ {\cellcolor{mygray1}{116.51}}} & {\cellcolor{mygray1}{67.33}} & {\cellcolor{mygray1}{34.43}} & {\cellcolor{mygray1}{20.25}} & {\cellcolor{mygray1}{15.11}} & {\cellcolor{mygray1}{12.80}} & \multicolumn{1}{|r}{{\cellcolor{mygray1}{20.54}}} & {\cellcolor{mygray1}{12.25}} & {\cellcolor{mygray1}{7.78}} & {\cellcolor{mygray1}{5.67}} & {\cellcolor{mygray1}{4.94}} & {\cellcolor{mygray1}{4.86}}\\
& \multicolumn{1}{|r}{{\cellcolor{mygray3}{PLMSS}}} & \multicolumn{1}{|r}{{\cellcolor{mygray3}{ - }}}&{\cellcolor{mygray3}{ - }}&{\cellcolor{mygray3}{ - }}&{\cellcolor{mygray3}{ - }}&{\cellcolor{mygray3}{ - }}&{\cellcolor{mygray3}{ - }}&\multicolumn{1}{|r}{{\cellcolor{mygray3}{ - }}}&{\cellcolor{mygray3}{ - }}&{\cellcolor{mygray3}{ - }}&{\cellcolor{mygray3}{ - }}&{\cellcolor{mygray3}{ - }}&{\cellcolor{mygray3}{ - }}\\ \hline 
\multirow{3}{*}{Asc/Desc} & \multicolumn{1}{|r}{{\cellcolor{mygray2}{TTK}}} & \multicolumn{1}{|r}{ {\cellcolor{mygray2}{442.15}}} & {\cellcolor{mygray2}{269.68}} & {\cellcolor{mygray2}{197.04}} & {\cellcolor{mygray2}{167.68}} & {\cellcolor{mygray2}{158.19}} & {\cellcolor{mygray2}{152.42}} & \multicolumn{1}{|r}{{\cellcolor{mygray2}{35.12}}} & {\cellcolor{mygray2}{28.23}} & {\cellcolor{mygray2}{24.54}} & {\cellcolor{mygray2}{22.92}} & {\cellcolor{mygray2}{22.38}} & {\cellcolor{mygray2}{22.63}}\\
& \multicolumn{1}{|r}{{\cellcolor{mygray1}{MSCEER}}} & \multicolumn{1}{|r}{ {\cellcolor{mygray1}{167.60}}} & {\cellcolor{mygray1}{99.48}} & {\cellcolor{mygray1}{53.12}} & {\cellcolor{mygray1}{28.71}} & {\cellcolor{mygray1}{15.89}} & {\cellcolor{mygray1}{11.63}} & \multicolumn{1}{|r}{{\cellcolor{mygray1}{34.77}}} & {\cellcolor{mygray1}{24.07}} & {\cellcolor{mygray1}{21.35}} & {\cellcolor{mygray1}{19.88}} & {\cellcolor{mygray1}{18.97}} & {\cellcolor{mygray1}{19.25}}\\
& \multicolumn{1}{|r}{{\cellcolor{mygray3}{PLMSS}}} & \multicolumn{1}{|r}{ {\cellcolor{mygray3}{39.61}}} & {\cellcolor{mygray3}{20.08}} & {\cellcolor{mygray3}{10.64}} & {\cellcolor{mygray3}{5.73}} & {\cellcolor{mygray3}{3.07}} & {\cellcolor{mygray3}{2.31}} & \multicolumn{1}{|r}{{\cellcolor{mygray3}{4.40}}} & {\cellcolor{mygray3}{2.23}} & {\cellcolor{mygray3}{1.16}} & {\cellcolor{mygray3}{0.68}} & {\cellcolor{mygray3}{0.43}} & {\cellcolor{mygray3}{0.36}}\\ \hline
\multirow{3}{*}{MS} & \multicolumn{1}{|r}{{\cellcolor{mygray2}{TTK}}} & \multicolumn{1}{|r}{ {\cellcolor{mygray2}{4.78}}} & {\cellcolor{mygray2}{2.44}} & {\cellcolor{mygray2}{1.40}} & {\cellcolor{mygray2}{1.02}} & {\cellcolor{mygray2}{0.73}} & {\cellcolor{mygray2}{0.66}} & \multicolumn{1}{|r}{{\cellcolor{mygray2}{0.46}}} & {\cellcolor{mygray2}{0.23}} & {\cellcolor{mygray2}{0.18}} & {\cellcolor{mygray2}{0.16}} & {\cellcolor{mygray2}{0.11}} & {\cellcolor{mygray2}{0.10}}\\
& \multicolumn{1}{|r}{{\cellcolor{mygray1}{MSCEER}}} & \multicolumn{1}{|r}{{\cellcolor{mygray1}{ - }}}&{\cellcolor{mygray1}{ - }}&{\cellcolor{mygray1}{ - }}&{\cellcolor{mygray1}{ - }}&{\cellcolor{mygray1}{ - }}&{\cellcolor{mygray1}{ - }}&\multicolumn{1}{|r}{{\cellcolor{mygray1}{ - }}}&{\cellcolor{mygray1}{ - }}&{\cellcolor{mygray1}{ - }}&{\cellcolor{mygray1}{ - }}&{\cellcolor{mygray1}{ - }}&{\cellcolor{mygray1}{ - }}\\
& \multicolumn{1}{|r}{{\cellcolor{mygray3}{PLMSS}}} & \multicolumn{1}{|r}{ {\cellcolor{mygray3}{0.32}}} & {\cellcolor{mygray3}{0.16}} & {\cellcolor{mygray3}{0.09}} & {\cellcolor{mygray3}{0.05}} & {\cellcolor{mygray3}{0.03}} & {\cellcolor{mygray3}{0.02}} & \multicolumn{1}{|r}{{\cellcolor{mygray3}{0.04}}} & {\cellcolor{mygray3}{0.02}} & {\cellcolor{mygray3}{0.01}} & {\cellcolor{mygray3}{0.01}} & {\cellcolor{mygray3}{0.00}} & {\cellcolor{mygray3}{0.00}}\\ \hline
\multirow{3}{*}{Index} & \multicolumn{1}{|r}{{\cellcolor{mygray2}{TTK}}} & \multicolumn{1}{|r}{ {\cellcolor{mygray2}{152.24}} }& {\cellcolor{mygray2}{90.30}} & {\cellcolor{mygray2}{46.73}} & {\cellcolor{mygray2}{24.59}} & {\cellcolor{mygray2}{12.80}} & {\cellcolor{mygray2}{9.43}} & \multicolumn{1}{|r}{{\cellcolor{mygray2}{26.73}}} & {\cellcolor{mygray2}{16.49}} & {\cellcolor{mygray2}{8.83}} & {\cellcolor{mygray2}{4.58}} & {\cellcolor{mygray2}{2.56}} & {\cellcolor{mygray2}{2.07}}\\
& \multicolumn{1}{|r}{{\cellcolor{mygray1}{{\cellcolor{mygray1}{MSCEER}}}}} & \multicolumn{1}{|r}{{\cellcolor{mygray1}{117.00}}} & {\cellcolor{mygray1}{58.92}} & {\cellcolor{mygray1}{30.87}} & {\cellcolor{mygray1}{16.20}} & {\cellcolor{mygray1}{8.68}} & {\cellcolor{mygray1}{6.24}} & \multicolumn{1}{|r}{{\cellcolor{mygray1}{15.70}}} & {\cellcolor{mygray1}{8.37}} & {\cellcolor{mygray1}{4.90}} & {\cellcolor{mygray1}{3.00}} & {\cellcolor{mygray1}{2.07}} & {\cellcolor{mygray1}{1.77}}\\
& \multicolumn{1}{|r}{{\cellcolor{mygray3}{PLMSS}}} & \multicolumn{1}{|r}{ {\cellcolor{mygray3}{39.00}}} & {\cellcolor{mygray3}{19.62}} & {\cellcolor{mygray3}{10.24}} & {\cellcolor{mygray3}{5.28}} & {\cellcolor{mygray3}{2.72}} & {\cellcolor{mygray3}{1.89}} & \multicolumn{1}{|r}{{\cellcolor{mygray3}{4.81}}} & {\cellcolor{mygray3}{2.42}} & {\cellcolor{mygray3}{1.27}} & {\cellcolor{mygray3}{0.67}} & {\cellcolor{mygray3}{0.41}} & {\cellcolor{mygray3}{0.29}}\\ \hline
\multirow{3}{*}{Geometry} & \multicolumn{1}{|r}{{\cellcolor{mygray2}{TTK}}} & \multicolumn{1}{|r}{ {\cellcolor{mygray2}{491.21}}} & {\cellcolor{mygray2}{263.75}} & {\cellcolor{mygray2}{143.22}} & {\cellcolor{mygray2}{83.25}} & {\cellcolor{mygray2}{50.67}} & {\cellcolor{mygray2}{41.04}} & \multicolumn{1}{|r}{{\cellcolor{mygray2}{94.11}}} & {\cellcolor{mygray2}{49.84}} & {\cellcolor{mygray2}{27.69}} & {\cellcolor{mygray2}{16.05}} & {\cellcolor{mygray2}{10.02}} & {\cellcolor{mygray2}{8.25}}\\
& \multicolumn{1}{|r}{{\cellcolor{mygray1}{MSCEER}}} & \multicolumn{1}{|r}{{\cellcolor{mygray1}{ - }}}&{\cellcolor{mygray1}{ - }}&{\cellcolor{mygray1}{ - }}&{\cellcolor{mygray1}{ - }}&{\cellcolor{mygray1}{ - }}&{\cellcolor{mygray1}{ - }}&\multicolumn{1}{|r}{{\cellcolor{mygray1}{ - }}}&{\cellcolor{mygray1}{ - }}&{\cellcolor{mygray1}{ - }}&{\cellcolor{mygray1}{ - }}&{\cellcolor{mygray1}{ - }}& {\cellcolor{mygray1}{ - }}\\
& \multicolumn{1}{|r}{{\cellcolor{mygray3}{PLMSS}}} & \multicolumn{1}{|r}{ {\cellcolor{mygray3}{8.99}}} & {\cellcolor{mygray3}{6.68}} & {\cellcolor{mygray3}{4.12}} & {\cellcolor{mygray3}{2.81}} & {\cellcolor{mygray3}{2.03}} & {\cellcolor{mygray3}{1.79}} & \multicolumn{1}{|r}{{\cellcolor{mygray3}{0.59}}} & {\cellcolor{mygray3}{0.42}} & {\cellcolor{mygray3}{0.27}} & {\cellcolor{mygray3}{0.18}} & {\cellcolor{mygray3}{0.13}} & {\cellcolor{mygray3}{0.14}}\\ \hline  \hline
DGF + & \multicolumn{1}{|r}{{\cellcolor{mygray2}{TTK}}} & \multicolumn{1}{|r}{ {\cellcolor{mygray2}{1,921.16}}} & {\cellcolor{mygray2}{1,045.49}} & {\cellcolor{mygray2}{608.77}} & {\cellcolor{mygray2}{392.42}} & {\cellcolor{mygray2}{283.09}} & {\cellcolor{mygray2}{\textbf{246.39}}} & \multicolumn{1}{|r}{{\cellcolor{mygray2}{214.50}}} & {\cellcolor{mygray2}{122.71}} & {\cellcolor{mygray2}{75.82}} & {\cellcolor{mygray2}{50.81}} & {\cellcolor{mygray2}{38.26}} & {\cellcolor{mygray2}{\textbf{34.78}}}\\
Index + & \multicolumn{1}{|r}{{\cellcolor{mygray1}{MSCEER}}} & \multicolumn{1}{|r}{ {\cellcolor{mygray1}{401.12}}} & {\cellcolor{mygray1}{225.73}} & {\cellcolor{mygray1}{118.43}} & {\cellcolor{mygray1}{65.16}} & {\cellcolor{mygray1}{39.68}} & {\cellcolor{mygray1}{\textbf{30.67}}} & \multicolumn{1}{|r}{{\cellcolor{mygray1}{71.02}}} & {\cellcolor{mygray1}{44.70}} & {\cellcolor{mygray1}{34.03}} & {\cellcolor{mygray1}{28.56}} & {\cellcolor{mygray1}{25.98}} & {\cellcolor{mygray1}{\textbf{25.87}}}\\
Asc/Des & \multicolumn{1}{|r}{{\cellcolor{mygray3}{PLMSS}}} & \multicolumn{1}{|r}{ {\cellcolor{mygray3}{78.61}}} & {\cellcolor{mygray3}{39.70}} & {\cellcolor{mygray3}{20.89}} & {\cellcolor{mygray3}{11.01}} & {\cellcolor{mygray3}{5.79}} & {\cellcolor{mygray3}{\textbf{4.20}}} & \multicolumn{1}{|r}{{\cellcolor{mygray3}{9.21}}} & {\cellcolor{mygray3}{4.65}} & {\cellcolor{mygray3}{2.43}} & {\cellcolor{mygray3}{1.35}} & {\cellcolor{mygray3}{0.84}} & {\cellcolor{mygray3}{\textbf{0.65}}}\\ \hline
    \end{tttabular} \hfill%
    \label{tab:strong_scaling}
\end{table*}

\begin{figure*}[!htb]
\includegraphics[width=0.95\textwidth]{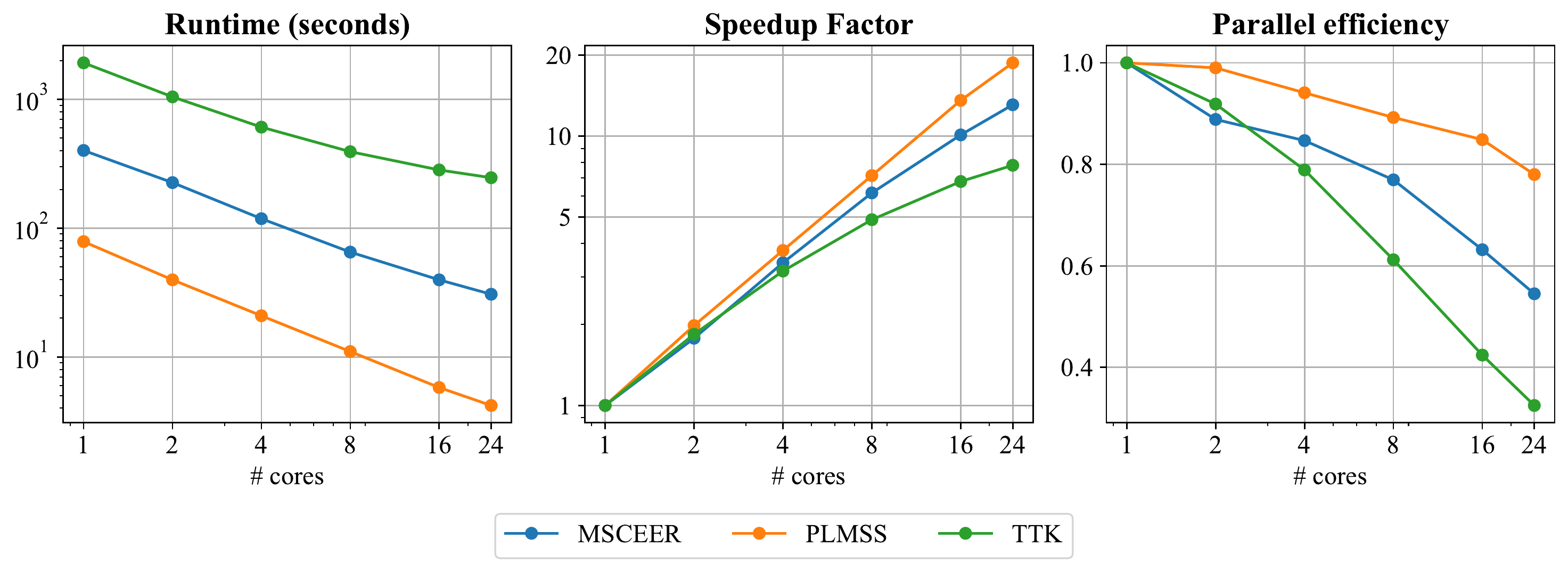}
\caption{Rayleigh-Taylor instability dataset~\cite{miranda} performance of all three algorithms at a resolution of  $512^3$. The left graph shows the runtime sum of the DGF, ascending and descending segmentation, and indexing as a log-log plot regarding the number of cores used for the experiment. In the middle, the speedup factor, i.e. the runtime of 1 thread divided by the runtime of $x$ threads, is also shown as a log-log plot regarding core counts. On the right, the parallel efficiency, i.e. the speedup factor divided by the number of cores, is represented in a semi-log plot.\label{fig:miranda_performance}}
\end{figure*}

\begin{figure*}[!htb]
\includegraphics[width=0.95\textwidth]{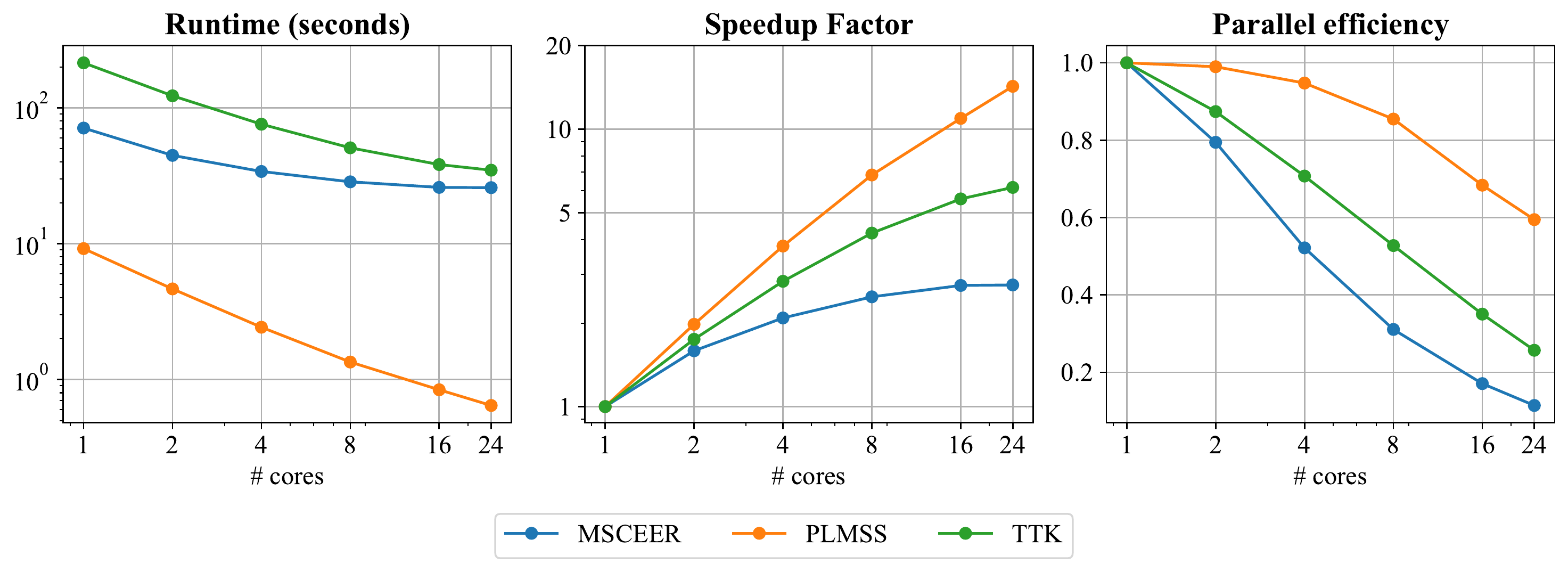}
\caption{Foot dataset~\cite{ttkData} performance of all three algorithms at a resolution of  $256^3$. The left graph shows the runtime sum of the DGF, ascending and descending segmentation, and indexing as a log-log plot regarding the number of cores used for the experiment. In the middle, the speedup factor, i.e. the runtime of 1 thread divided by the runtime of $x$ threads, is also shown as a log-log plot of core counts. On the right, the parallel efficiency, i.e. the speedup factor divided by the number of cores, is represented in a semi-log plot.\label{fig:foot_performance}}
\end{figure*}

In this section, the computational performance of our implementation is analyzed and compared to the MS complex implementation of TTK\cite{ttk17} and MSCEER\cite{msceer}. As hinted by the authors of MSCEER, we use the ``steepest\_lstar'' and ``extractms'' packages, as they supply the fastest implementation without accurate geometry. Both strong scaling studies show that the PLMSS is scaling well with core count due to the mentioned improvements. We utilize the Rayleigh-Taylor instability (Miranda) dataset~\cite{miranda} at a resolution of $512^3$ and the Foot dataset~\cite{ttkData} at a resolution of $256^3$ for the computation speed comparisons of all three algorithms.

\subsection{Algorithmic improvements}
In general, three main aspects of the PLMSS result in a strongly reduced computation time as compared to the MS complex. First, the segmentation of the domain is improved by path compression, which is much faster than computing a discrete gradient field. Second, the multi-label marching tetrahedra algorithm supports computing the separating geometries in a well-scaling way. Third,  splitting the marching tetrahedra algorithms into indexing and geometry creation steps allows the allocation of the resources needed in the geometry creation step without additional computations.  

\subsubsection{Segmentation}
The segmentation of the domain into the ascending and descending manifold and the intersection of both manifolds, called MS manifold, are computed differently for the MS complex and the PLMSS. Both MS complex implementations require a discrete gradient field to be computed first, then following the v-paths along the gradient field to assign labels to each vertex. However, the PLMSS segmentation utilizes path compression to assign each vertex to its designated minimum or maximum, without the need for any additional structure other than the order field.

For path compression to work, all neighbors of each vertex must be visited once to get the largest and smallest neighbor of that vertex. With this information, the maximum can be found iteratively by assigning its largest neighbor's largest neighbor to the vertex. This process is executed in multiple iterations, where each iteration finds the designated maximum of a vertex or a vertex closer to the designated maximum. Here, each time the step length is doubled, yielding extremum assignment in $log(s)$ steps for each vertex, where $s$ is the number of vertices on the integral line of the vertex to the extremum. Also, the iterations get smaller every time, as more and more vertices are assigned to their extremum.

\autoref{tab:strong_scaling} shows the timings of those steps in the first three rows, where "DGF" refers to the discrete gradient field computation, "Asc/Desc" refers to the ascending and descending segmentation, and "MS" refers to the MS segmentation. Please note that the MSCEER algorithm does not compute an MS segmentation in the provided implementation.

Even in a single-threaded environment, the performance gains of retrieving the ascending and descending segmentations already show strong improvements in the computation time of the PLMSS compared to the MS complex implementations. For the Miranda dataset, PLMSS only needs a total of 39.61s for the computation of the ascending and descending segmentation. Comparing this to the MS complex implementations (MSCEER: 284.11s / TTK: 1768.91s) leaves us at a speedup of 7x and 44x respectively.

\subsubsection{Multi-Label Marching Tetrahedra}
After segmenting the domain into various regions, region-separating geometries can be created that help to visualize the segmentation effectively. Again, the PLMSS and the MS complex implementations differ in the realization of this step. Regarding the MS complex implementations, paths between critical point pairs have to be traced on the DGF, whereas the PLMSS uses a marching tetrahedra algorithm.

Marching tetrahedra algorithms scale very well as they are executed per vertex. In our implementation, the binary code, as described in \autoref{sec:marchingtet}, and the number of triangles created per thread are computed in a preliminary step. This allows the allocation of the exact amount of memory needed for the triangles. In a follow-up step, this enables direct writing of triangles to memory.

\autoref{tab:strong_scaling} shows the timings of those steps in the 4th and 5th row, where "Index" refers to the computation of simplex indices that will spawn triangles, and "Geometry" represents writing the triangles to memory. Please note that the MSCEER algorithm does not compute the geometry itself in the provided implementation, but only gathers the indices of relevant simplices.

A comparison of the single thread timings with the Miranda dataset shows that the indexing is almost four times faster compared to TTK and three times faster than MSCEER. These speedups come from the excessive use of lookup tables and a per tetrahedra execution that does not require any tracing of paths. With the additional memory counting in the indexing step and the lower number of triangles created, the geometry-creating part strongly improved regarding computation times.

\subsection{TTK vs. MSCEER}
Both MS complex solutions are slightly different in the specifics of their implementation. First of all, TTK uses a simplicial complex that uses tetrahedra and triangles as top-level simplices, whereas MSCEER uses cubes and squares. This already leads to roughly $324\%$ more total cells and six times more top-level cells in the case of the TTK-based implementation, such as the MS complex or PLMSS. This will have an effect on the runtime of the algorithms, as well as the resolution of the extracted region-separating geometries. Still, this is only a limitation of the current TTK version and might change in the future to trade accuracy for runtime efficiency.

Another issue, when comparing both algorithms with each other, is the output generated by each implementation. The version provided by Gyulassy et al.\cite{msceer} only provides an ascending and descending segmentation of the domain and the indices of top-level cells that would spawn region-separating geometries. The MS complex representation and the surface geometries would be created by a different software at runtime that was not supplied with the library.

\subsection{Strong Scaling Setup}
Both strong scaling studies were carried out on a dual Intel XEON SP 6126 node with 24 CPU cores and 384GB of RAM. For each algorithm, various timings were created, starting with the computation of the discrete gradient field, ascending and descending segmentation, MS segmentation, indexing of tetrahedra or cubes that generate region-separating geometries, and writing the region-separating geometries to memory. The results for both studies are shown in \autoref{tab:strong_scaling}, where the last row shows the total time to get an ascending and descending segmentation and mark all top-level cells that generate region-separating geometries. To mimic the layout of the 2-cells of the MS complex, the region separators of the PLMSS were computed for those results.

For each combination of the dataset, the number of threads, and the algorithm, the experiment was executed $10$ times. From these 10 runs, the best and the worst ones, regarding total computation time, were discarded. The remaining 8 runs were averaged to achieve more stable results.

\subsection{Strong Scaling: Rayleigh-Taylor Instability}
Utilizing the Rayleigh-Taylor instability dataset~\cite{miranda}, a strong scaling analysis was carried out for a $512^3$ subset of the data, simplified with a persistence threshold of $0.1$. Computation times, speedup factor (i.e. the time a single thread takes divided by the time the current number of threads take), and parallel efficiency (i.e. speedup factor divided by the number of threads) are plotted against the number of cores, shown in \autoref{fig:miranda_performance}.

The total execution time in the last row of \autoref{tab:strong_scaling} shows that the PLMSS is more than an order of magnitude faster than TTK and 5 to 7 times faster than MSCEER. As the PLMSS still has to be executed on roughly three times the cells, this is still an improvement of an order of magnitude regarding the time per cell. The speedup factor and parallel efficiency also show a clear trend that PLMSS is scaling very well with more cores. In this regard, MSCEER beats TTK in terms of scalability but clearly performs worse against PLMSS. The speedup factor graph also hints that more cores would be beneficial in future experiments as PLMSS still scales well at 24 cores. Due to insufficient computational resources, we were unable to offer a larger strong scaling analysis.

\subsection{Strong Scaling: CT Scan of a Human Foot}
The foot dataset~\cite{ttkData} was chosen to represent smaller data sizes with a resolution of $256^3$, simplified with a persistence threshold of $110$. It consists of a CT scan of the tip of a foot, where the threshold of 110 was chosen to represent each bone with its own region.

Regarding the total execution time at $24$ Threads in \autoref{tab:strong_scaling}, a speedup of over 50x and almost 40x (TTK and MSCEER) can be achieved using PLMSS. The resulting graphs in \autoref{fig:foot_performance} also show clear improvements regarding parallel efficiency, as PLMSS ($0,59$) still achieves good results that hint towards using even more cores, where TTK ($0,26$) and MSCEER ($0,11$) are already in a range where more cores do not strongly improve runtime performance and communication overhead takes over.

\subsection{Discussion}
For both datasets, the PLMSS showed good improvements over the MSCEER and TTK implementation, yielding a great parallel efficiency. Especially, for the foot dataset, great runtime performance improvements of roughly 40x were achieved. Even with the 5-7x runtime improvement regarding the Rayleigh-Taylor instability dataset, 6x more top-level cells had to be traversed compared to MSCEER.

\section{Limitations}
\label{sec:Limitations}
Our entire approach aims at efficiently computing ascending and descending segmentations of the input scalar field.
Its output is not a complete Morse-Smale complex.
First, it does not capture saddle-saddle connectors, which may be useful in certain applications.
Second, it does not output an explicit CW complex modeling the Morse-Smale complex (i.e. where vertices encode critical points, 1-dimensional cells encode separatrices, 2-dimensional cells encode separating geometries, and 3-dimensional cells encode regions with identical integration extremities), only the domain segmentation is provided. 
This can be detrimental in applications involving post-processing of the Morse-Smale complex, such as regular remeshing~\cite{DongBGPH06} or hierarchical simplification~\cite{GyulassyNPBH06}.
For these applications, a standard algorithm based on discrete Morse theory (such as the one available in TTK~\cite{ttk17} or MSCEER~\cite{msceer}) should be preferred.

\section{Conclusion and Future Work}
\label{sec:Conclusion}
The presented algorithm describes a well-scaling approach to computing MS segmentations, allowing for speedups of more than an order of magnitude compared to two MS complex implementations. Utilizing path compression to create the segmentations allows us to quickly extract the labels for the multi-label marching tetrahedra algorithm that powers the generation of region-separating geometries. The algorithm is not only faster but also has a lower memory footprint, as no discrete gradient vector field and little preprocessing of the triangulation is needed. Only a scalar field saving the 5-bit index representation for each vertex has to be created. The utilization of only top-level cells and vertices  simplifies triangulation generation. Additionally, maxima at the border of the MS complex are not allowed by DMT design, often leading to missing border regions in each segmentation. This issue is fixed by segmenting the whole domain, also retrieving all border maxima in the process. Regarding the generated separating geometries, several use cases have been presented that show the applicability of our approach where the MS complex failed to deliver without expensive post-processing using saddle-saddle 2-cell cancellation. Simply speaking, our algorithm allows us to extract areas of influence of minima, maxima, and minima-maxima pairs by separating their boundaries with two available separating geometries that can be triggered by three segmentation options. Still, this does not invalidate the Morse-Smale complex as we only compute a segmentation and not the complex itself. Features like the saddle-saddle connectors are not computed. Therefore, we conclude that the PLMSS is a versatile tool to generate MS segmentations in a well-scaling parallel way, allowing users to explore their data much faster, while still being able to fall back to the MS complex on demand. 

For future work, we are planning to improve the PLMSS in various ways. The load on each of the threads can be imbalanced when a particular thread receives a lot of triangles to generate. Here, a workload balance system could be introduced. To scale to even larger datasets, an MPI implementation will be provided to enable distributed memory execution. The knowledge gained from creating the segmentations will be implemented into the TTK MS complex implementation to improve its performance and useability. As shown in \autoref{sec:performance}, a thorough comparison between MS complex implementations is out of the scope of this paper. Comparing all publicly available implementations and characterizing them in terms of input simplicial complex, handling of functions that are not Morse, available simplification models (pre- vs. post-simplification), output options, and memory footprint would be beneficial. Additionally, a version of the PLMSS using voxels as top-level cells in 3D could potentially speed up the computation considerably and would downsize the memory footprint even more. Also, the effect of path compression might be applicable to the computation of the MS complex, so additional research in integrating it might be of interest to achieve better runtime efficiency of MS complex implementations.

\ifCLASSOPTIONcompsoc
  \section*{Acknowledgments}
    This research was funded by the Deutsche Forschungsgemeinschaft (DFG, German Research Foundation) – 252408385 – IRTG 2057.
    This work is also partially supported by the European Commission grant ERC-2019-COG \emph{``TORI''} (ref. 863464, \url{https://erc-tori.github.io/}).
\else
  \section*{Acknowledgment}
    This research was funded by the Deutsche Forschungsgemeinschaft (DFG, German Research Foundation) – 252408385 – IRTG 2057.
    This work is also partially supported by the European Commission grant ERC-2019-COG \emph{``TORI''} (ref. 863464, \url{https://erc-tori.github.io/}).
\fi



\bibliographystyle{IEEEtran}
\bibliography{paper}

%

\begin{IEEEbiography}[{\includegraphics[width=1in,height=1.25in,clip,keepaspectratio]{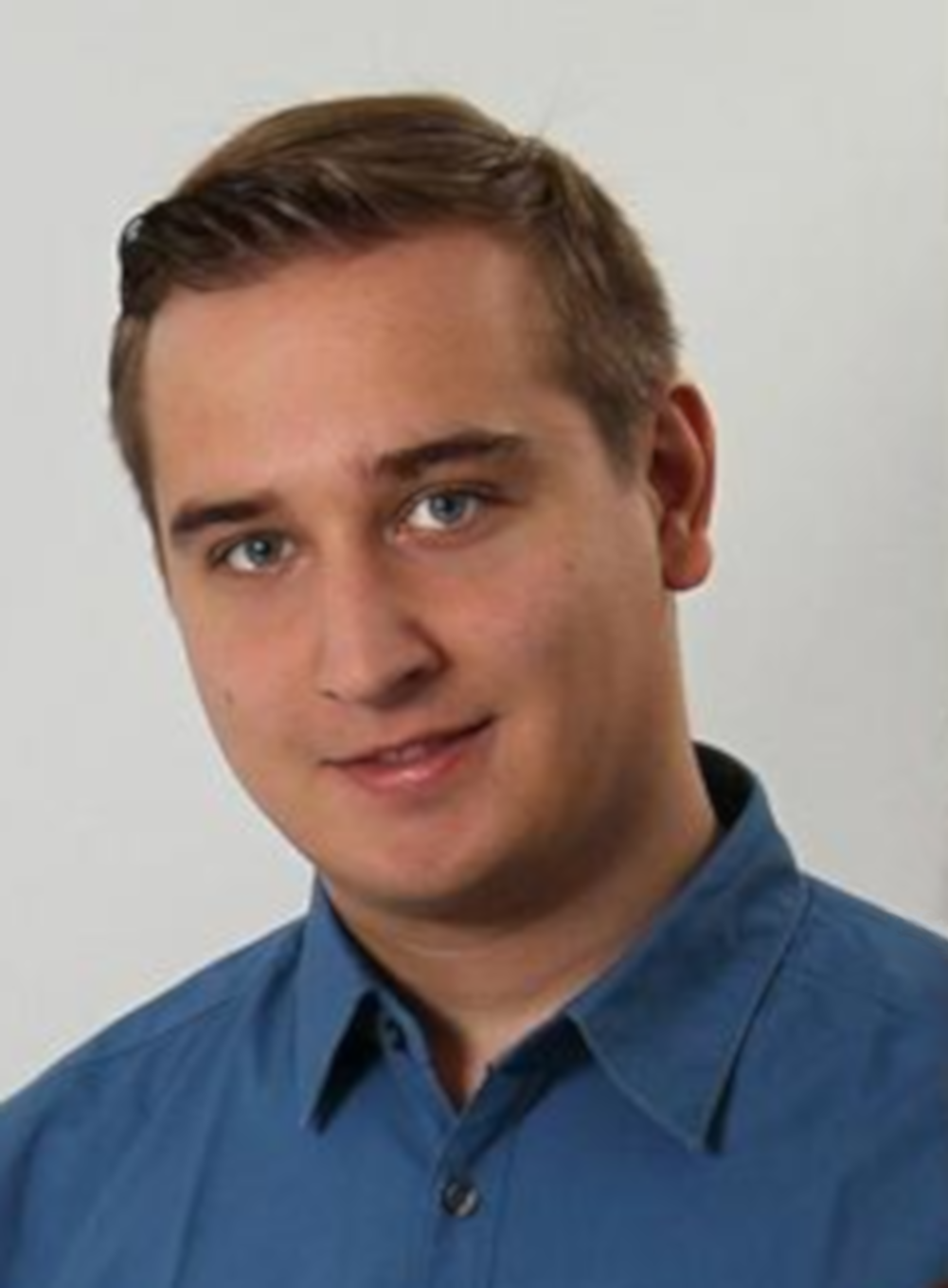}}]{Robin G. C. Maack}
received the Master's degree in computer science in October 2020 from the University of Kaiserslautern. He started as a student assistant in January 2017 and now further develops his projects as a PhD student. His research interests include topological data analysis, medical image analysis and visualization, biochemical visualization, and uncertainty visualization.
\end{IEEEbiography}

\begin{IEEEbiography}[{\includegraphics[width=1in,height=1.25in,clip,keepaspectratio]{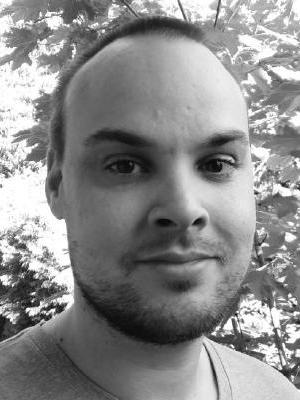}}]{Jonas Lukasczyk}
received his Ph.D. degree from the Visual Information Analysis Group, Technische Universitat Kaiserslautern, Germany, where he also studied applied computer science and mathematics. His recent work focuses on topology-based characterization of features and their evolution in large-scale simulations. 
\end{IEEEbiography}

\begin{IEEEbiography}[{\includegraphics[width=1in,height=1.25in,clip,keepaspectratio]{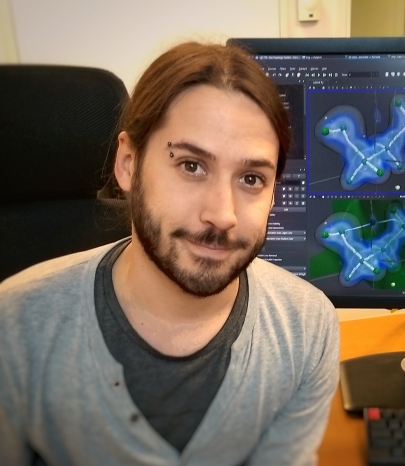}}]{Julien Tierny}
received the PhD degree in computer science from the University of Lille, in
2008. He is currently a CNRS senior scientist, affiliated with
Sorbonne University. Prior to his
CNRS tenure, he held a Fulbright fellowship (U.S.
Department of State) and was a postdoctoral
researcher at the Scientific Computing and Imaging Institute at the University of Utah. His research expertise lies in topological
methods for data analysis and visualization. He is the founder and lead
developer of the Topology ToolKit (TTK), an open source library for topological data analysis.
\end{IEEEbiography}

\begin{IEEEbiography}[{\includegraphics[width=1in,height=1.25in,clip,keepaspectratio]{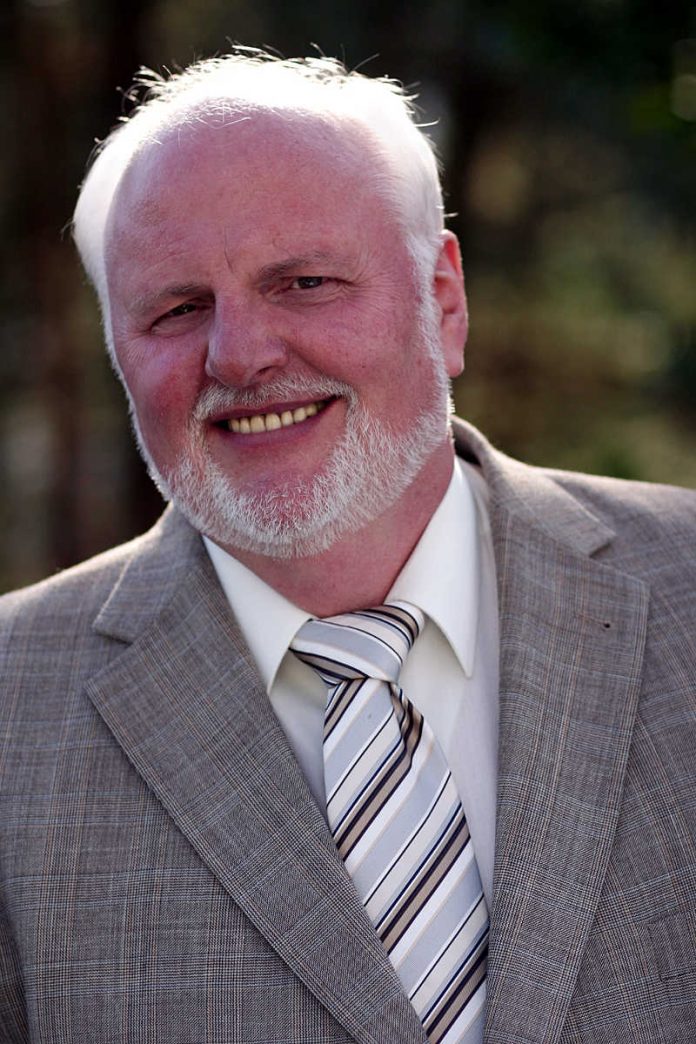}}]{Hans Hagen}
is a computer science professor at University
of Kaiserslautern and an adjunct professor at University of California,
Davis. He received a Bachelor’s degree in computer science, a Master
degree in mathematics from the University of Freiburg and a PhD in
mathematics (geometry) from the University of Dortmund. His main
research interests are scientific visualization and geometric modeling.
He is a member of the IEEE Visualization Academy of Science, and
he got the IEEE Visualization Career Award, the ACM Solid Modeling
Pioneer Award and the John Gregory Memorial Award among others.
\end{IEEEbiography}

\begin{IEEEbiography}[{\includegraphics[width=1in,height=1.25in,clip,keepaspectratio]{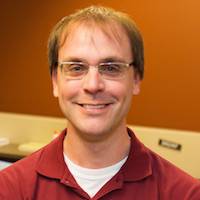}}]{Ross Maciejewski}
is a professor with the School of Computing and Augmented Intelligence at Arizona State University and director of the Center for Accelerating Operational Efficiciency - a Department of Homeland Security Center of Excellence. His primary research interests include the areas of geographical visualization and visual analytics.
\end{IEEEbiography}

\begin{IEEEbiography}[{\includegraphics[width=1in,height=1.25in,clip,keepaspectratio]{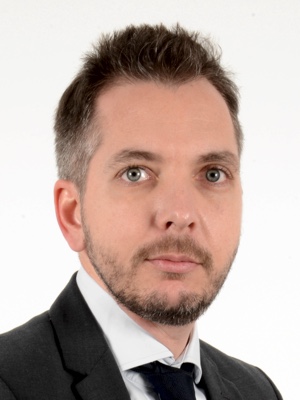}}]{Christoph Garth}
received the PhD degree in
computer science from Technische Universität
(TU) Kaiserslautern in 2007. After four years as
a postdoctoral researcher with the University of
California, Davis, he rejoined TU Kaiserslautern
where he is currently a full professor of computer
science. His research interests include largescale data analysis and visualization, in situ visualization, topology-based methods in visualization, and interdisciplinary applications of visualization.
\end{IEEEbiography}




\end{document}